\documentclass[twocolumn]{aastex62}
\usepackage{hyperref}

\usepackage{xcolor}
\usepackage{rotating}
\usepackage{amsmath}

\shorttitle{Core Mass Function in Massive Protoclusters}
\shortauthors{O'Neill et al.}

\newcommand{\bmin}{$M_{\rm{min}} $ }

\begin{document}

\title{The Core Mass Function Across Galactic Environments. III.\\Massive Protoclusters}

\author{Theo J. O'Neill}
\affiliation{Dept. of Astronomy, University of Virginia, Charlottesville, Virginia 22904, USA}

\author{Giuliana Cosentino}
\affiliation{Dept. of Space, Earth \& Environment, Chalmers University of Technology, Gothenburg, Sweden}

\author{Jonathan C. Tan}
\affiliation{Dept. of Astronomy, University of Virginia, Charlottesville, Virginia 22904, USA}
\affiliation{Dept. of Space, Earth \& Environment, Chalmers University of Technology, Gothenburg, Sweden}

\author{Yu Cheng}
\affiliation{Dept. of Astronomy, University of Virginia, Charlottesville, Virginia 22904, USA}

\author{Mengyao Liu}
\affiliation{Dept. of Astronomy, University of Virginia, Charlottesville, Virginia 22904, USA}

\begin{abstract}
The stellar initial mass function (IMF) is fundamental for many areas of astrophysics, but its origin remains poorly understood. It may be inherited from the core mass function (CMF) or arise as a result of more chaotic, competitive accretion. Dense, gravitationally bound cores are seen in molecular clouds and some observations have suggested that the CMF is similar in shape to the IMF, though translated to higher masses by a factor of $\sim3$. Here we measure the CMF in 28 dense clumps within 3.5~kpc that are likely to be central regions of massive protoclusters, observed via $1.3\:{\rm{mm}}$ dust continuum emission by the ALMAGAL project. We identify 222 cores using the dendrogram algorithm with masses ranging from 0.04 to $252\:M_{\odot}$. We apply completeness corrections for flux and number recovery, estimated from core insertion and recovery experiments. At higher masses, the final derived CMF is well described by a single power law of the form $dN/d\:{\textrm{log}}\:M\propto\:M^{-\alpha}$ with $\alpha\simeq0.94\pm0.08$. However, we find evidence of a break in this power-law behavior between $\sim5$ and $15\:M_{\odot}$, which is, to our knowledge, the first time such a break has been found in distant ($\gtrsim 1$~kpc) regions by ALMA. We compare this massive protocluster CMF with those derived using the same methods in the G286 protocluster and a sample of Infrared Dark Clouds. The massive protocluster CMF is significantly different, i.e., containing more massive cores, which is a potential indication of the role of environment on the CMF and IMF.
\end{abstract}

\keywords{stars: formation -- ISM: clouds}

\section{Introduction}\label{S:intro}
Understanding the origin of the stellar initial mass function (IMF) is crucial for many areas of astrophysics, from the evolution of stellar clusters to the structure and chemical composition of galaxies.  Stars are known to form in cold, dense, gravitationally-bound cores within molecular clouds, the masses of which may be related to the resultant stellar masses. Therefore, characterizing the core mass function (CMF) is likely to be of fundamental importance for explaining the stellar IMF and its connection to the environmental conditions of molecular clouds.

The IMF appears to be universal \citep[see reviews by, e.g.,][]{bastian_universal_2010,kroupa_stellar_2013}, though with some evidence for variations in extreme environments. The IMF has been shown to have a peak at the low-mass end, i.e., at $M_{\rm *,peak}\sim 0.5\:M_\odot$, and follow a power law at the high-mass end of the form:

\begin{equation}  
\frac{d N}{d\: {\rm log}\: M}\propto{M}^{-\alpha}.
\label{eq:pl}
\end{equation}

\citet{salpeter_luminosity_1955} derived $\alpha\simeq1.35$ for the mass range between 0.4 and 10~$M_{\odot}$, which continues to be consistent with the results of more recent studies.
Observations of cores in nearby regions, i.e., $\lesssim 1$~kpc, have shown that the CMF may approximate the shape of the IMF, with values of $\alpha \gtrsim 1.0$ to $1.5$ \citep[e.g.,][]{alves_mass_2007,2010A&A...518L.102A,konyves_census_2015}.  

The existence of a peak in the IMF suggests that the distribution is not scale-free, which would imply the existence of some physical process that results in a characteristic mass of new stars.  If the IMF is directly determined by the CMF, observing a similar peak in the CMF would allow the efficiency of star formation to be derived from this shift in mass scale. Evidence for a peak or ``break'' in the CMF has also occasionally been observed, with the break shifted to higher masses than the peak of the IMF by a factor of $\sim 3$ \citep[e.g.,][]{alves_mass_2007}.

Characterizing the CMF in more distant ($\gtrsim 1$ kpc) star-forming regions has proven to be challenging due to the need for higher angular resolution interferometric observations, but is critical for a robust comparison of the CMF to the IMF as nearby star-forming regions do not produce significant numbers of massive ($\gtrsim 8\:M_{\odot}$) stars. 
\citet{motte_unexpectedly_2018} found $\alpha = 0.90\pm0.06$ for masses $M\gtrsim 1.6\:M_\odot$ from a sample of 105 cores in the massive cloud W43-MM1 ($d\sim 5.5$ kpc), as identified with the {\it getsources} algorithm plus a manual pruning of elongated structures.  \citet{sanhueza_alma_2019} derived $\alpha = 1.07 \pm 0.09$ for $M \geq 0.6 \ M_{\odot}$ in a sample of 294 cores identified using the dendrogram algorithm in 12 IRDC clumps ($d \sim 2.9 \ - 5.4$ kpc) observed at 1.3 mm by ALMA.  Similarly, \citet{lu_alma_2020} derived $\alpha = (0.83 - 1.07)$ for $M \geq 5.9 \  M_{\odot}$ from a sample of $\sim 800$ cores identified using the dendrogram algorithm in four massive molecular clouds in the Central Molecular Zone ($d \sim 8.2$ kpc) observed by ALMA.

In the first paper in this series we measured the CMF from 1.3~mm ALMA continuum observations of the G286 protocluster ($d=2.5$ kpc) finding $\alpha=1.24\pm0.17$ over the mass range $M\gtrsim 1\:M_\odot$, based on 76 cores identified by the fiducial dendrogram algorithm (see below) \citep[Paper I]{cheng_core_2018}. In the second paper, we analyzed a sample of IRDC clumps ($d =$2.4 - 5.7 kpc) finding $\alpha=0.86\pm0.11$ over the same mass range, based on 107 cores found using the same identification methods \citep[Paper II]{liu_core_2018}.  In these two previous studies, corrections were made for flux (mass) recovery and number recovery based on artificial core insertion experiments. 

Here, using the same methods as Papers I and II, we conduct an analysis of 1.3~mm continuum survey data of 28 dense, relatively nearby ($d < 3.5$ kpc) clumps observed by the ALMA Cycle 7 project ALMAGAL.  In \S\ref{S:obs} we describe the observational data and analysis methods. In \S\ref{S:results} we present our results on the derived core population and the CMF, including corrections for completeness, a comparison of power-law fitting methods, an analysis of the evidence for there being a break in the power law, a discussion of log-normal fits, and a discussion of the spatial distribution of the cores. We make direct comparison of our CMF results with those of Papers I and II in \S\ref{S:discussion}. We discuss the implications of our results and conclude in \S\ref{S:conclusion}.

\section{Observations and Analysis Methods}\label{S:obs}
\subsection{Source Selection}
\begin{deluxetable*}{cccccccccccc}
\tabletypesize{\scriptsize}
\tablecaption{Overview of Clump Sample\label{tab:sources} }
\tablehead{
  \colhead{ID} & \colhead{ALMA ID} &  \colhead{Hi-GAL ID} & \colhead{$\ell$}  & \colhead{$b$} & \colhead{$d$} & \colhead{$R_{\rm cl}$} & \colhead{$M_{\rm cl}$} & \colhead{$\Sigma_{\rm cl}$} & \colhead{Type} & \colhead{RMS Noise} & \colhead{$N_{c}$} \\
 \colhead{}  &  \colhead{} &  \colhead{} & \colhead{($\degr$)} &  \colhead{($\degr$)} &  \colhead{(kpc)} & \colhead{(pc)} & \colhead{($M_{\odot}$)} & \colhead{(g cm$^{-2}$)} & \colhead{} &  \colhead{(mJy $\rm beam^{-1}$)} & \colhead{}    \\
 \vspace{-0.4cm}
 }
\startdata
 G18.30 &              82445 &      9393 &   18.30268 & -0.38638 &      2.96 &    0.18 &   946 &       1.87 &     2 &  0.39 &          5 \\  G24.52 &             109432 &     12565 &   24.52526 & -0.13900 &      3.21 &    0.12 &   674 &       3.22 &     1 &  0.20 &          5 \\ G294.51 &  G294.5117-01.6205 &     30535 &  294.51124 & -1.62099 &      2.20 &    0.05 &   111 &       2.91 &     2 &  0.33 &          6 \\ G305.13 &             704454 &     33862 &  305.13755 &  0.06798 &      3.50 &    0.15 &  1135 &       3.42 &     2 &  0.41 &         10 \\ G309.98 &  G309.9796+00.5496 &     36003 &  309.97979 &  0.54944 &      3.47 &    0.15 &   329 &       0.98 &     2 &  0.30 &          5 \\ G309.99 &             721509 &     36008 &  309.99155 &  0.51362 &      3.47 &    0.21 &  1246 &       1.82 &     2 &  0.32 &          2 \\ G314.21 &             737579 &     38106 &  314.21835 &  0.25370 &      3.44 &    0.16 &   153 &       0.39 &     2 &  0.23 &          8 \\ G314.22 &             737588 &     38107 &  314.22068 &  0.27165 &      3.44 &    0.06 &   334 &       5.98 &     2 &  0.51 &         15 \\ G316.58 &  G316.5871-00.8086 &     38894 &  316.58689 & -0.80883 &      3.21 &    0.09 &   217 &       1.69 &     2 &  0.24 &          6 \\ G316.80 &             744714 &     39000 &  316.79965 & -0.05594 &      2.70 &    0.15 &   824 &       2.55 &     2 &  1.05 &         12 \\ G316.81 &             744757 &     39010 &  316.81128 & -0.05753 &      2.70 &    0.16 &   948 &       2.45 &     2 &  0.85 &          3 \\ G317.40 &             746985 &     39290 &  317.40833 &  0.11021 &      2.82 &    0.13 &   864 &       3.18 &     2 &  0.45 &          9 \\ G319.86 &             754973 &     40323 &  319.86682 &  0.78652 &      2.91 &    0.18 &   699 &       1.48 &     2 &  0.22 &         12 \\ G320.22 &             756911 &     40575 &  320.22739 &  0.87267 &      2.83 &    0.17 &   510 &       1.20 &     2 &  0.18 &          5 \\ G320.24 &             757085 &     40595 &  320.24635 & -0.29425 &      0.76 &    0.03 &    26 &       1.59 &     2 &  0.31 &          8 \\ G326.34 &             776981 &     43273 &  326.34137 &  0.50437 &      3.01 &    0.10 &   219 &       1.56 &     2 &  0.27 &          4 \\ G327.39 &             783304 &     44067 &  327.39265 &  0.19949 &      3.42 &    0.09 &   494 &       3.82 &     2 &  0.28 &         14 \\ G341.92 &             853658 &     52705 &  341.92906 & -0.16936 &      3.50 &    0.24 &  3021 &       3.61 &     2 &  0.25 &          6 \\ G341.94 &             853716 &     52713 &  341.94232 & -0.16600 &      3.50 &    0.13 &  1390 &       5.35 &     2 &  0.47 &         14 \\ G343.23 &             858727 &     53332 &  343.23903 & -0.71313 &      3.19 &    0.23 &   558 &       0.68 &     2 &  0.20 &          3 \\ G343.52 &  G343.5213-00.5171 &     53468 &  343.52148 & -0.51743 &      2.00 &    0.06 &   151 &       2.37 &     2 &  0.30 &          4 \\ G343.90 &  G343.9033-00.6713 &     53639 &  343.90376 & -0.67103 &      2.00 &    0.06 &    91 &       1.56 &     2 &  0.24 &          2 \\ G344.06 &             861762 &     53713 &  344.06344 & -0.65038 &      2.70 &    0.21 &   930 &       1.35 &     2 &  0.21 &          6 \\ G344.72 &             864399 &     54029 &  344.72568 & -0.54141 &      2.54 &    0.16 &   526 &       1.42 &     1 &  0.18 &          4 \\ G345.18 &             866226 &     54283 &  345.18157 &  1.04660 &      1.93 &    0.08 &   604 &       6.70 &     2 &  0.52 &          9 \\ G345.33 &             867071 &     54405 &  345.33634 &  1.02052 &      1.94 &    0.11 &   721 &       3.76 &     2 &  0.41 &          9 \\ G345.50 &  G345.5043+00.3480 &     54527 &  345.50430 &  0.34808 &      2.16 &    0.05 &   468 &      10.67 &     2 &  1.38 &         33 \\ G346.35 &             871581 &     55038 &  346.35546 &  0.10629 &      3.50 &    0.14 &   446 &       1.48 &     2 &  0.13 &          3
\enddata
\tablecomments{$N_{c}$ is the number of dendrogram-identified cores in the clump.  The radius, mass, $\Sigma_{\rm cl}$, and type were derived by \citet{elia_hi-gal_2017}, where type is the evolutionary classification of the clump in which 1 represents a prestellar clump and 2 represents a protostellar clump.}
\end{deluxetable*}

We selected sources from the ALMA Cycle 7 project ``ALMAGAL: ALMA Evolutionary Study of High Mass Protocluster Formation in the Galaxy'' (PI: S. Molinari). ALMAGAL was designed to observe continuum and line emission near $1.4\:$mm from $\gtrsim$1000 high-mass Galactic clumps ($d<$~7.5~kpc and $M_{\rm cl}> 500\:M_{\odot}$) identified by the {\it Herschel} Hi-GAL survey \citep{elia_hi-gal_2017}.

\begin{figure*}
    \centering
    \includegraphics[width=0.95\textwidth]{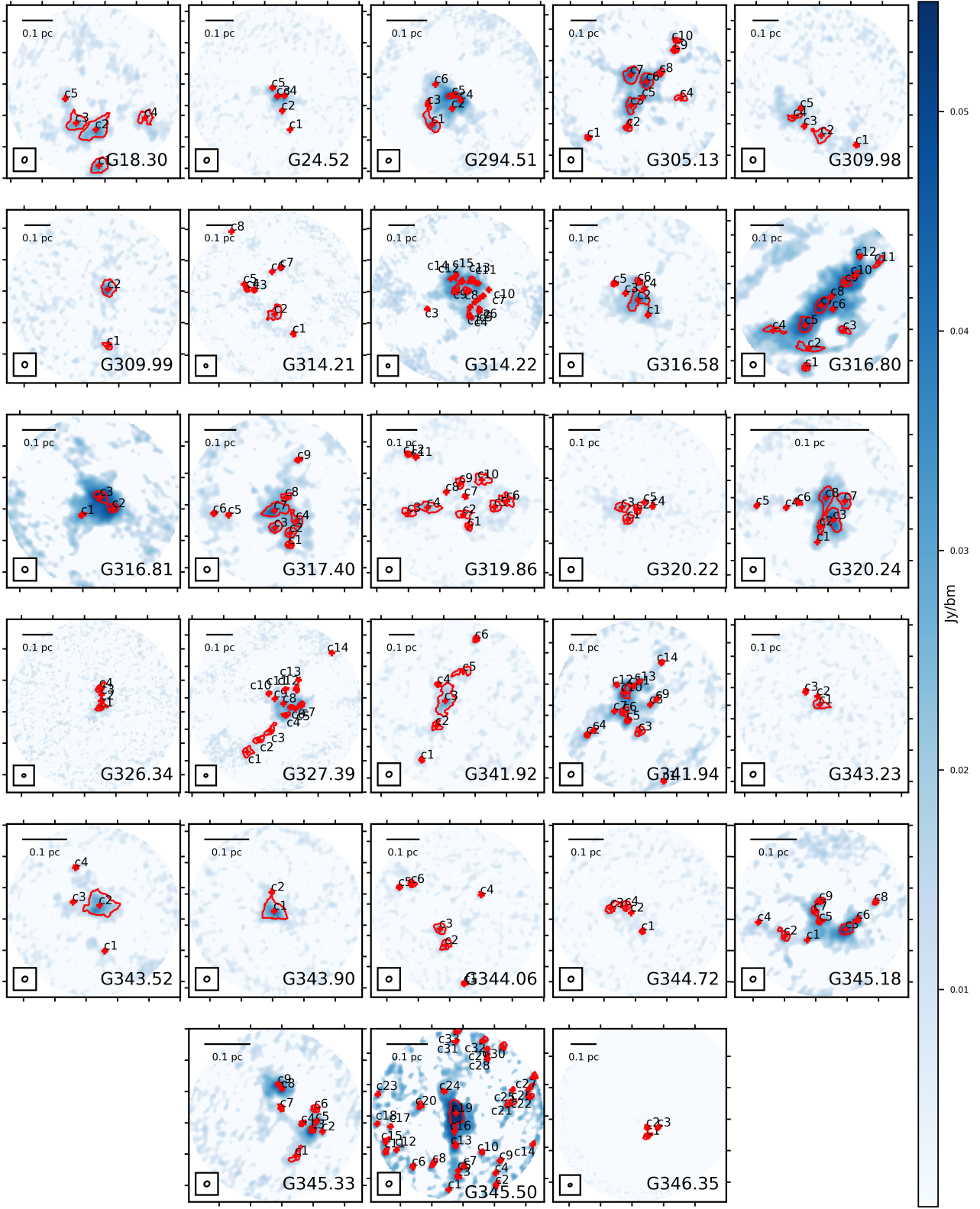}
    \caption{1.37 mm continuum images of our sample of 28 dense clumps observed by the ALMAGAL survey.  The synthesized beam is shown in the bottom left corner of each panel.  The cores identified by the dendrogram algorithm are marked, with red contours showing ``leaf'' structures.  Ticks on edges of images are spaced 7.2\arcsec\ apart. }
    \label{fig:ALMAGALsources}
\end{figure*}

To be able to probe down to relatively low core masses,  i.e., $\sim1\:M_\odot$, we limited our analysis to sources within 3.5~kpc that had 12m-array observations delivered to the {\it ALMA} archive by May 2020. This yields a sample of 28 sources, which are shown in Fig.~\ref{fig:ALMAGALsources} and described in Table~\ref{tab:sources}.

We note that this is not a complete sample of all such massive clumps within 3.5 kpc of the Sun and is only a small fraction of the data that will eventually come from the ALMAGAL survey. However, we have no information to suggest that this sample is biased in any particular way, e.g., in terms of evolutionary stage or local environmental conditions. We thus consider that this clump sample is representative of conditions near the centers of massive protoclusters. As we will see, the sample is large enough that the number of cores identified in the clumps is more than double that of previous samples analyzed in this series of papers.

\subsection{Core Identification}
To identify cores, we followed the methods used in our studies of the G286 protocluster (Paper I) and IRDCs (Paper II), thus allowing a direct comparison of the derived CMFs. We used the dendrogram algorithm \citep{rosolowsky_structural_2008} as implemented in the python package {\tt astrodendro} to identify cores.  

We estimated the 1$\sigma$ root mean square (RMS) noise within each pointing by finding the mean value of $\sigma$ in 50 randomly-placed synthesized-beam-sized-areas of the pre-primary-beam corrected continuum images with a maximum intensity less than 3.5 mJy $\textrm{beam}^{-1}$.  The median value of $\sigma$ was 0.30 mJy $\textrm{beam}^{-1}$.  We identified three clumps which had significantly higher noise levels: G316.80 ($\sigma = 1.05$ mJy $\textrm{beam}^{-1}$), 316.81 ($\sigma = 0.85$ mJy $\textrm{beam}^{-1}$), and G345.50 ($\sigma = 1.38$ mJy $\textrm{beam}^{-1}$).  We attribute these elevated ``noise'' estimates to the presence of cleaning residuals, perhaps including effects from real emission features that are just outside the primary beam.

We required parent structures (trunks) in the dendrogram to have a minimum intensity $F_{\rm min}$ of $4\sigma$. We also required substructures (\textit{branches} and \textit{leaves}) to have a minimum increase in intensity $\delta$ of $1\sigma$ and a minimum area $S_{\rm{min}}$ of half the synthesized beam to be considered independent entities.  We defined cores as the leaves of the dendrogram.   
These parameter choices are the same as used in Papers I and II; however, we diverge from Paper I's methods and follow those used in Paper II by identifying cores in images that have not been corrected for the primary beam.  The sample of IRDCs studied in Paper II is similar to this sample of massive clumps in that its data consist of multiple individual pointings, while Paper I focused on a mosaic of only one region.  A consequence of this difference is that, while images in Paper I possessed a relatively uniform noise level, the noise within the images in Paper II and in this work vary as a function of position.  Attempting to identify cores in primary-beam-corrected images would be compromised as a result of this effect.     

We note that the effects of using the {\tt clumpfind} algorithm \citep{williams_determining_1994} instead of the dendrogram method were explored in Paper I, and we do not repeat this analysis here.  We instead concentrate on comparing the fiducial method CMFs obtained from the sources we study here and those derived from the G286 protocluster and IRDC clump environments. We explore the effects of the choice of dendrogram parameters in \S\ref{S:dendro}.

\subsection{Core Mass Estimation}
We estimated masses of cores by assuming optically thin thermal emission from dust, as in Papers I and II. The total mass surface density corresponding to a given specific intensity of mm continuum emission is
\begin{eqnarray}
\Sigma_{\rm mm} & = & 0.369 \frac{F_\nu}{\rm mJy}\frac{(1\arcsec)^2}{\Omega} \frac{\lambda_{1.3}^3}{\kappa_{\nu,0.00638}}
 \nonumber\\
 & & \times  \left[{\rm exp}\left(0.553 T_{d,20}^{-1}
  \lambda_{1.3}^{-1}\right)-1\right]\:{\rm g\:cm^{-2}}\\
 &\rightarrow & 0.272 \frac{F_\nu}{\rm mJy}\frac{(1\arcsec)^2}{\Omega}\:{\rm g\:cm^{-2}}\nonumber,
\label{eq:Sigmamm}
\end{eqnarray}
where $F_{\nu}$ is the total integrated flux over solid angle
$\Omega$, $\kappa_{\nu,0.00638}\equiv\kappa_\nu/({\rm
  6.38\times10^{-3}\:cm^2\:g}^{-1})$ is the dust absorption
coefficient, $\lambda_{1.3}=\lambda/1.30\:{\rm mm}$ and
$T_{d,20}=T_d/20\:{\rm K}$ with $T_d$ being the dust temperature. To
obtain the above fiducial normalization of $\kappa_\nu$, we assumed an
opacity per unit dust mass $\kappa_{\rm 1.3mm,d}=0.899\: {\rm cm^2
 \ g}^{-1}$ (i.e., the value from the moderately coagulated thin ice mantle model of \citealt{ossenkopf_dust_1994}), which results in $\kappa_{\rm 1.3mm}= {\rm
  6.38\times10^{-3}\:cm^2\:g}^{-1}$ using a
gas-to-refractory-component-dust ratio of 141 \citep{draine_physics_2011}. The numerical factor following the $\rightarrow$ in the final line shows the fiducial case where $\lambda_{1.3}=1$ and $T_{d,20}=1$.

Core masses can then be calculated as 
\begin{eqnarray}  
M & = & \Sigma_{\rm mm} A = 0.113 \frac{\Sigma_{\rm mm}}{\rm g\:cm^{-2}} \frac{\Omega}{(1\arcsec)^2} \left(\frac{d}{\rm 1\:kpc}\right)^2  \:M_\odot\\
 & \rightarrow & 0.192 \frac{F_\nu}{\rm mJy} \left(\frac{d}{\rm 2.5\:kpc}\right)^2  \:M_\odot,\nonumber
\label{eq:coremass} 
\end{eqnarray}
where $A$ is the projected area of the core, $d$ is the source
distance, and the final evaluation is for fiducial temperature
assumptions of 20~K. In the absence of measurements of temperature for each source, we assumed dust temperatures $T_{d} =$ 20~K. We note that {\it Herschel}-derived dust temperature maps have much lower angular resolution ($\gtrsim 30\arcsec$) than the sizes of the ALMA-identified cores. For sources close to the Galactic plane, they are also subject to significant uncertainties depending on if and how background subtraction is performed \citep{lim_2016}. Finally, we expect most of the detect cores to be protostellar sources that are internally heated and have a range of temperatures that will be different, generally higher, than that of the surrounding ambient clump. 

If temperatures of 15~K or 30~K were to be adopted, the mass estimates would differ by factors of 1.48 and 0.604, respectively.  We note that if systematic temperature variations exist with increasing flux, i.e., if brighter cores are warmer, then the masses of the higher-mass cores could be overestimated. However, with the current data that are available, it is difficult to estimate how significant this effect might be. Higher resolution estimates of temperatures, e.g., from $\rm NH_3$ or from multiwavelength images across the peak of the spectral energy distribution (SED), are needed to improve this situation.  We also note that individual masses depend on distance, with mass proportional to distance squared, so a $10\%$ error in the Hi-GAL distance estimate to a particular clump would yield a $20\%$ error in mass, which would systematically affect all the cores in the clump. However, for the global sample of 28 clumps we do not expect distance uncertainties to lead to systematic over- or underestimation of masses.

\subsection{Core Flux Recovery and Completeness Corrections}\label{S:recovery}
\begin{figure*}
    \centering
    \includegraphics[width=0.9\textwidth]{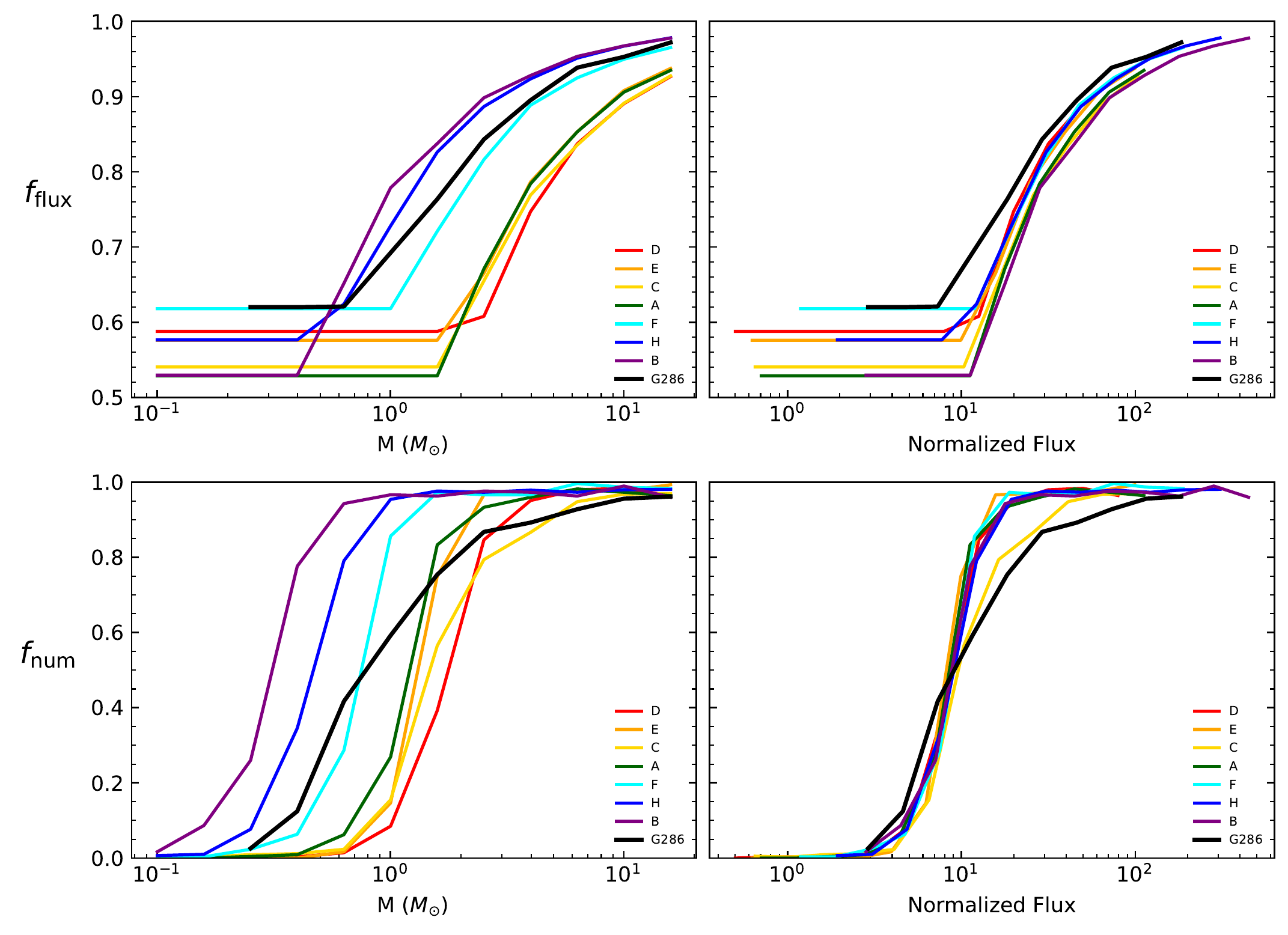}
    \caption{\textit{Top left:} Median flux recovery fractions $f_{\rm{flux}}$ versus core mass $\textit{M}$ for the dendrogram core finding method as derived from the G286 protocluster studied in Paper I and the seven IRDCs (A-H, ordered in legend by distance) studied in Paper II.  \textit{Top right:} Median flux recovery fractions $f_{\rm{flux}}$ versus the flux of the inserted cores normalized by the RMS noise in the region into which it was inserted.  \textit{Bottom left:} Mean value of the number recovery fraction $f_{\rm{num}}$ versus core mass $\textit{M}$ for the dendrogram core finding method.  \textit{Bottom right:} Median number recovery fractions $f_{\rm{num}}$ versus the flux of the inserted cores normalized by the RMS noise in the region into which it was inserted.}
    \label{fig:corr_curves}
\end{figure*}

An accurate comparison of the CMF and IMF is dependent on the
completeness of the sample of cores.  Underestimating the flux, and by extension
mass, of individual cores or failing to identify all cores in a region can significantly
bias the derived CMF.  In Papers I and II, we derived correction factors to estimate a complete, i.e., ``true'', CMF from an observed ``raw'' CMF.  Artificial cores of varying masses were randomly inserted into real clump images to estimate the amount of the
flux typically detected by the dendrogram algorithm as a function of inserted mass. This information was used to
calculate a flux recovery fraction $f_{\rm flux}$, defined as the median value of the ratio of recovered to true flux, and create a flux-corrected CMF from the raw CMF. A number recovery fraction $f_{\rm num}$ of the inserted cores of a given mass that were successfully identified through the fiducial methods was also calculated, and used to create a final ``true'' CMF from the flux-corrected CMF.  

The above correction factors were first calculated as a function of inserted core mass. We then converted these correction factors to be a function of the ``true'' flux of the inserted core (using eq.~\ref{eq:Sigmamm} and eq.~\ref{eq:coremass}), and then a function of the ``true'' flux of the inserted core normalized by the RMS noise $\sigma$ of the region into which the core was inserted.  We present $f_{\rm{flux}}$ and $f_{\rm{num}}$ as functions of inserted mass and of normalized flux in Fig. \ref{fig:corr_curves}.  Although the correction factors vary between regions when considered as a function of inserted mass, they closely approximate each other when considered as a function of normalized flux.  This similarity in functional form suggests that these factors may be applicable to a variety of Galactic clump source morphologies, such as those exhibited in our ALMAGAL clump sample.  We attribute the slight variations in these correction curves to imperfect measurements of $\sigma$ and varying degrees of crowding in some regions, which is likely especially pronounced in the G286 protocluster.  We adopted the median values of these factors as universal values with which to correct our observed ``raw'' CMFs from our entire clump sample.  

\section{Results}\label{S:results}
\subsection{Continuum Images and Average Core Properties}

The continuum images of the 28 massive clumps are shown in Fig. \ref{fig:ALMAGALsources} with the boundaries of the cores (i.e., leaves) identified by the dendrogram algorithm.  We found 222 cores in total and their properties are listed in Table \ref{tab:cores}.  Cores are named as, e.g., G18.30c1, G18.30c2, etc., in the region G18.30, with the numbering order ranging from lower to higher Galactic latitude.  The most populated clump is G345.50 with $N_c=33$ cores and all other clumps contain $N_c \leq 15$ cores each. 

``Raw'' masses ranged from 0.04 to 252 $M_{\odot}$ with a median mass of 3.0 $M_{\odot}$. The median radius of the cores was $R_{c} \sim 0.01$ pc ($R_{c} = \sqrt{A/\pi}$), where $A$ is the core's projected area).  We evaluated the mean mass surface density of the cores, $\Sigma_{c} = M/A$. We find a median values of $\Sigma_c =1.5$ g cm$^{-2}$.
We applied $f_{\rm{flux}}$ corrections to each core, which increased all masses, especially at the lower-mass end, and resulted in a new median mass of 4.6 $M_{\odot}$ and a new median 
$\Sigma_c = 2.6$ g cm$^{-2}$.

\subsection{The Core Mass Function (CMF)}

\subsubsection{Construction of the CMF}\label{S:cmfcon}
\begin{figure*}
    \centering
    \includegraphics[width=\textwidth]{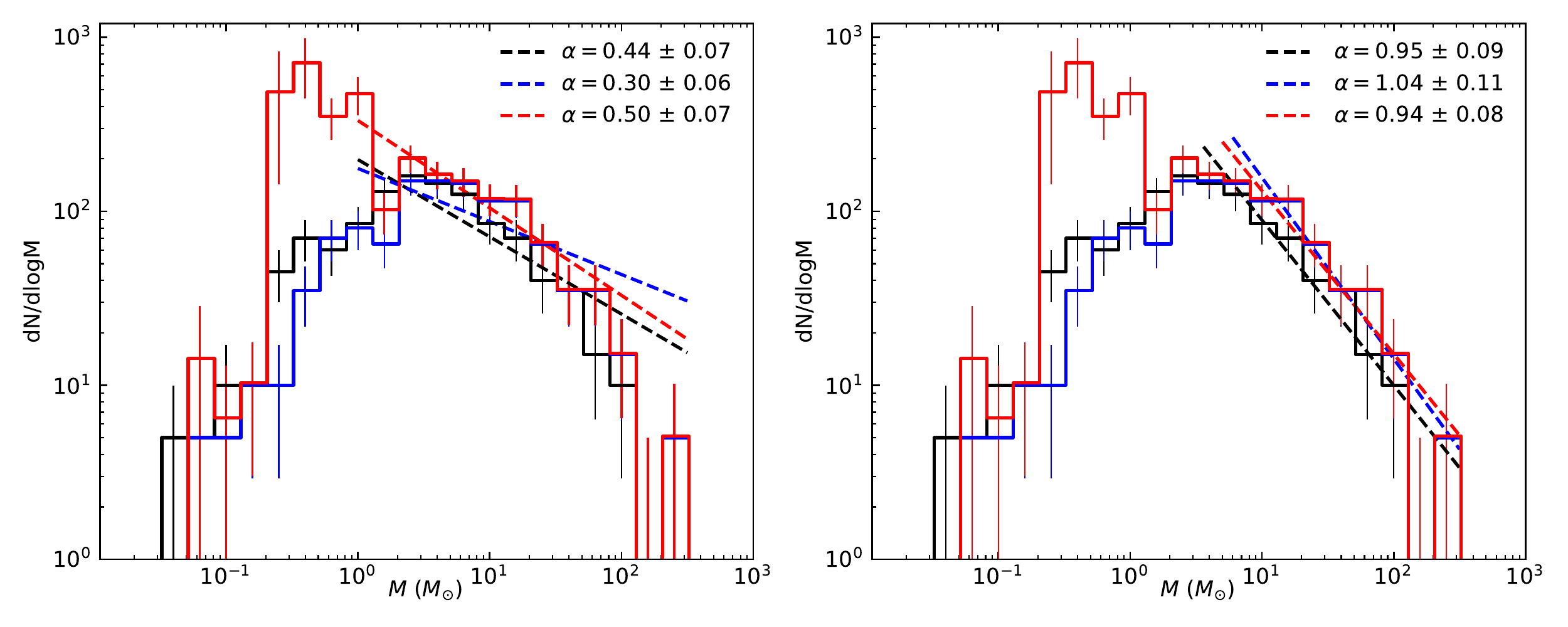}
    \caption{\textit{(a) Left:} The combined CMF of the 28 massive clump sample. The black histogram shows the original, ``raw'' CMF.  The blue histogram shows the CMF after flux correction and the red histogram shows the final, ``true'' CMF, i.e., after applying number recovery fraction correction to the flux-corrected CMF. The dashed black, blue and red lines show the best power law fits derived through weighted least squares fitting over the range from $M = 0.79 \ M_{\odot}$ up to the most massive populated bin. \textit{(b) Right:} The same as the left panel, but the dashed lines shown correspond to the best power-law fits derived through the fiducial MLE fitting process (see text), beginning at $M_{\rm{min}}$ $=3.6, 6.0$ and $5.0$ $M_{\odot}$, respectively.}
    \label{fig:almagal_fit_cmfs}
\end{figure*}


We created individual raw, flux-corrected, and number-corrected ``true'' CMFs for each clump. We then combined all sources together to create final CMFs.  All three final CMFs are shown in Fig. \ref{fig:almagal_fit_cmfs}, with the raw CMF in black, the flux-corrected CMF in blue, and the true CMF in red.  Masses were binned using the same methods as Papers I and II, with all bins equal in size in $\Delta \rm{log_{10}}M$ with five bins per dex, and a bin centered on 1 $M_{\odot}$.  We assumed $\textit{N}^{1/2}$ Poisson counting uncertainties for each bin.  Uncertainties on the final, ``true'' CMF are the same fractional size as those of the flux-corrected CMF's bins. However, note that these underestimate the true uncertainties, as they do not allow for additional uncertainty in $f_{\rm num}$.  

From inspection of Fig.~\ref{fig:almagal_fit_cmfs} we see that the raw CMF exhibits a broad peak between about 2 and 5 $M_{\odot}$, while for the flux-corrected CMF this peak is wider, extending from about 2 to 8 $M_{\odot}$. The number correction yields a significantly increased population of low mass ($\lesssim 1 M_{\odot}$) cores in the true CMF.  

\subsubsection{Single Power-Law Fits to the CMF}\label{S:fitting}

\begin{deluxetable}{c|c|cccc}
\centering
\tablecaption{Best-Fit Power-Laws\label{tab:plawparams}}
\tablehead{\colhead{CMF}&	\colhead{Type}&	\colhead{$M_{\rm min}$}&	\colhead{$\alpha$} &	\colhead{$\Delta\alpha$} & \colhead{$p$}}
\startdata  
Raw&	WLS &	0.79*&	0.44&	0.07&	$<$.001	\\
\nodata&	MLE&	0.79*&	0.55&	0.04&	$<$.001	\\
\nodata&	MLE-B &	0.79*&	0.67&	0.06&	$<$.001	\\
\nodata&	MLE&	3.6&	0.95&	0.09&	.15	\\
\hline						
Flux-corrected &	WLS&	0.79*&	0.30&	0.06&	$<$.001	\\
\nodata&	MLE&	0.79*&	0.48&	0.03&	$<$.001	\\
\nodata&	MLE-B &	0.79*&	0.56&	0.05&	$<$.001	\\
\nodata &	MLE&	6.0&	1.04&	0.11&	.11	\\
\hline						
True &	WLS&	0.79*&	0.50&	0.07&	$<$.001	\\
\nodata&	MLE-B &	0.79*&	0.60&	0.05&	$<$.001	\\
\nodata&	MLE-B &	5.0&	0.94&	0.08&	.12	
\enddata
\tablecomments{$M_{\rm{min}}$'s marked with a $*$ were assigned, rather than being found via the fitting process. The MLE type is a fit to the unbinned sample of cores, while MLE-B type is a fit to a binned histogram of the data.}
\end{deluxetable}

We fit a power law of the form of Equation (\ref{eq:pl}) to the binned raw, flux-corrected and true CMFs. First, we follow the fiducial weighted least squares (WLS) method of Papers I and II, which involves minimizing differences in the log of $d N / d {\rm log} M$, normalized using the average of the asymmetric Poisson errors. For the one empty bin at the high-mass end, we again follow Papers I and II and treat it as an effective upper limit by assuming that the point is a factor of 10 lower than the level if the bin had 1 data point and set the error bar such that it reaches up to the level if there were 1 data point in the bin. Note that the results are insensitive to these details of how the empty bin is treated.

The fiducial method of Papers I and II was to fit the CMF over a standard range from the mass bin centered at $1\:M_\odot$ (i.e., for cores down to a minimum mass $M_{\rm min} = 0.79\:M_\odot$) up to the highest mass bin that is populated. These fits are shown in Figure~\ref{fig:almagal_fit_cmfs}a and listed in Table~\ref{tab:plawparams}. We see that the values of the power law index, $\alpha$, are relatively low, e.g., $\alpha=0.50\pm 0.07$ for the true CMF, compared to the Salpeter index of 1.35. We also note that a single power law fit over this mass range is not a very accurate description of the CMFs, which is indicated by the fits having very low $p$ values ($<0.001$).

We also consider maximum likelihood estimation (MLE) methods for fitting a power-law to the CMF, which have been shown to outperform least-squares based methods \citep{clark_generalizations_1999,white_estimating_2008}. In particular, \citet{clauset_power-law_2009} proposed methods to reliably fit power-laws to distributions that are not binned into histograms through direct numerical maximization of the likelihood and this is the method we implement here for fits to the raw and flux-corrected CMFs. We note that several other CMF studies have also adopted aspects of these methods \citep[e.g.,][]{swift_discerning_2010,olmi_shape_2013,lu_alma_2020}. 

We assume a power-law form of $f(M)=C M^{-(\alpha+1)}= (\alpha / M_{\rm{min}}) (M/M_{\rm{min}})^{-(\alpha + 1)}$.  To estimate $\alpha$ for an unbinned sample of cores above some minimum mass $M_{\rm{min}}$ we applied an MLE \citep{newman_power_2005} in which 
\begin{equation}
    \alpha = n\left[ \sum_{i=1}^{n} \textrm{ln} \frac{M_{i}}{M_{\textrm{min}}} \right]^{-1},
\end{equation}
where $M_{i}$ are the masses of the cores that have masses $\geq M_{\rm{min}}$  and $n$ is the number of cores in the sample selected for the fit.  We first adopt $M_{\rm{min}}$ $=0.79$ $M_{\odot}$ (corresponding to the bin centered at $1\:M_{\odot}$). The standard error on $\alpha$ is $\Delta\alpha = \alpha / \sqrt{n}$.
We only fit to the raw and flux-corrected CMFs using this method, as the number-corrected CMF is defined by bin only.  
We also use a second MLE method (MLE-B) to fit to binned data, i.e., fitting power-laws directly to the histograms of the raw, flux-corrected, and true CMFs. This method is based on that proposed by \citet{virkar_power-law_2014}, with a log-likelihood function above some minimum bin $M_{\rm{min}}$ of the form:
\begin{equation}
    \mathcal{L} = n \  \alpha\ \textrm{ln}\ M_{\textrm{min}} + \sum_{i=\textrm{min}}^{k} h_{i}\ \textrm{ln}\ [M_{i}^{-\alpha}-M_{i+1}^{-\alpha}].
\end{equation}

The results of employing these MLE and MLE-B fitting methods with $M_{\rm{min}}$ $= 0.79 \ M_{\odot}$ are presented in Table \ref{tab:plawparams}.  We derived moderately steeper power-law indices than those found through WLS-based fitting, e.g., for the true CMF we derived $\alpha=0.60 \pm 0.05$ (via MLE-B), while the WLS method found $\alpha=0.50\pm0.07$. As illustrated for the raw and flux-corrected CMFs, the differences in the results of MLE and MLE-B fitting are at a similar level, with the latter yielding slightly steeper power law indices.

We next consider single power law fits to the CMF, but with the minimum mass of the fitting range, $M_{\rm min}$, allowed to vary to obtain the ``best'' description of the data, as described below. This analysis is important as a first step in identifying potential break points in the CMF, which may have a physical origin that is important for setting the characteristic masses of cores and, eventually, stars.

We first estimated the optimal $M_{\rm{min}}$ for each CMF through fitting power-laws beginning at every unique core mass (or mass bin) within the sample and calculating the Kolmogorov-Smirnov (KS) distance $D$ between each cumulative distribution and their corresponding fit:
\begin{equation}
    D = {\rm max}\left| F(M) - F_{n}(M)\right|,
\end{equation}
where $F(M)$ is the cumulative distribution function (CDF) of the power law model and $F_{n}(M)$ is the CDF of the sample. Identifying the mass $M_{\rm{min}}$ that minimizes a distance-based statistic provides an estimate of where power-law behavior is most likely to begin \citep[e.g.,][]{reiss_statistical_2007,clauset_power-law_2009}. 

However, the KS statistic is not especially sensitive to differences in the tails of distributions and it performs best when the compared distributions diverge near their medians.  This presents a particular problem when assessing asymmetrical distributions like power-laws. As an alternative distance metric that is more sensitive to differences in the tails of distributions, we also consider the Anderson-Darling statistic $A^{2}$:
\begin{equation}
    A^{2} = n \int_{-\infty}^{\infty} \frac{(F_n(M)-F(M))^2}{F(M)(1-F(M))} d F(M),
\end{equation}
where $n$ is the number of cores or bins included in the fit.  Other distance metrics such as Pearson's $\chi^{2}$ cumulative test statistic are also potentially applicable for this goal, but have been found to be less powerful than KS or AD based approaches \citep{virkar_power-law_2014}.

\begin{figure*}
    \centering
    \includegraphics[width=\textwidth]{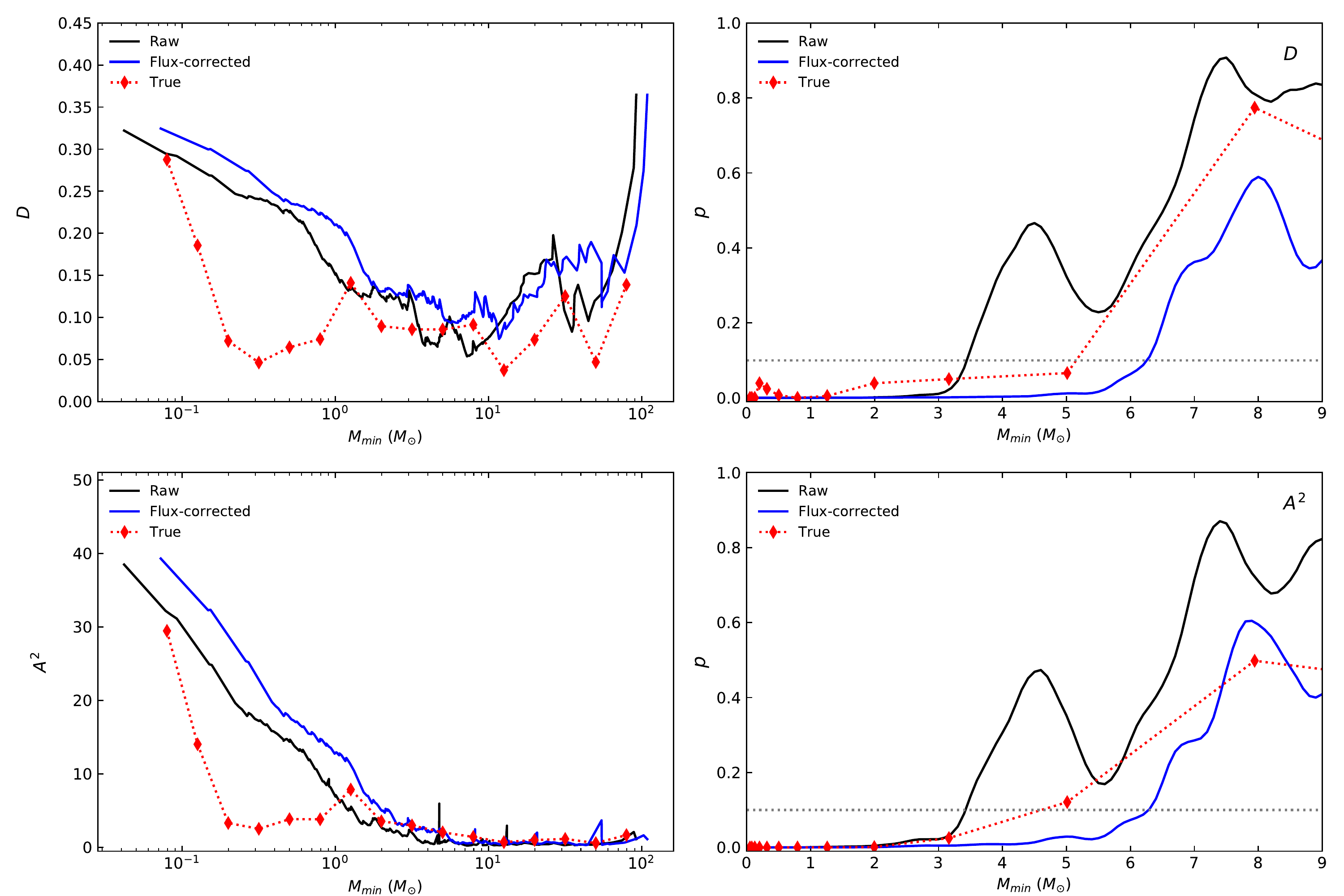}
    \caption{\textit{(a) Top left:} Kolmogorov-Smirnov statistic $D$ derived from best-fit power-law versus $M_{\rm{min}}$. The black curve is from fitting to the unbinned sample of raw core masses. The blue curve is from fitting to the unbinned sample of flux-corrected core masses. The red curve is from fitting to the binned number-corrected true CMF, with each bin marked by a diamond. 
    \textit{(b) Bottom left:} Anderson-Darling statistic $A^2$ derived from best-fit power-law versus $M_{\rm{min}}$. The black, blue and red curves have the same meaning as in (a). 
    \textit{(c) Top right:} The $p$-value of a given fit to the CMF as measured by the KS statistic, $D$, is compared to the fixed $M_{\rm{min}}$ from which the fit was derived. The $p$-value for each fit was generated from bootstrapped samples of size $N=3000$. The black, blue and red curves have the same meaning as in (a), but note the black and blue curves were smoothed using LOWESS regression (see text).  The black dotted line shows $p=.10$. 
    \textit{(d) Bottom right:} As (c), but now the $p$-value of a given fit to the CMF as measured by the AD statistic $A^2$ is compared to the fixed $M_{\rm{min}}$ from which the fit was derived.}
    \label{fig:d_mmins}  
\end{figure*}

We show the values of $D$ and $A^2$ as a function of $M_{\rm{min}}$ for the raw, flux-corrected, and true CMFs in Fig.~\ref{fig:d_mmins}.  There is a broad minimum of $D$ values between about 3 and $15\:M_{\odot}$ for the raw and flux-corrected CMFs, with absolute minima at $\sim 7$ and $\sim 12\:M_{\odot}$, respectively.  The true CMF shows a broad minimum between 0.2 and 50~$M_{\odot}$, with an absolute minimum of $D$ at 12.5~$M_{\odot}$.  In contrast, $A^2$ values decrease as $M_{\rm min}$ increases, and then plateau to near constant minimum values above $\sim3\:M_{\odot}$ for the raw and flux-corrected CMFs.  Values of $A^2$ for the true CMF remain roughly constant from about 8 to $80\:M_\odot$.
The absolute minimum values of $A^2$ for the raw, flux-corrected and true CMFs occur at 7, 8 and $50\:M_{\odot}$.  

These profiles of $D$ and $A^2$, which have broad ranges of $M_{\rm min}$ where the values are at a similar low level, indicate that a single power-law description of the CMF begins to become reasonable somewhere around $M_{\rm min}\sim3\:M_{\odot}$. To quantify further, we calculated the goodness of fit of a power-law as a function of $M_{\rm min}$. 
However, typical hypothesis testing is not applicable here, since if the parameters of a model are derived from the data to which the model is then compared, then the standard critical values used to generate $p$-values from the KS or AD statistics are no longer valid because the tests' requirements of independence are not met \citep{lilliefors_kolmogorov-smirnov_1967}. To circumvent this issue, correct $p$-values can instead be estimated via bootstrap resampling \citep{babu_goodness--fit_2004,clauset_power-law_2009}.

To carry out this analysis, we first fit a power-law to the observed CMF using the fiducial MLE method, but with a fixed $M_{\rm{min}}$.  We recorded the values of the KS statistic, $D$, and AD statistic, $A^2$, and counted how many cores were below $M_{\rm{min}}$, $n_{\rm{low}}$, and how many were above, $n_{\rm{high}}$. Next, we generated a synthetic CMF through a two-part, semi-parametric method based on the procedure proposed by \citet{clauset_power-law_2009}.  To create the low-mass end of the synthetic CMF ($M < M_{\rm{min}}$), we randomly selected $n_{\rm low}$ cores from the pool of the $n_{\rm low}$ real cores. Note, this selection was done with replacement, i.e., any core could be chosen more than once. This ensured that the low-mass end of the synthetic sample would follow roughly the same distribution as the real CMF (which is not guaranteed to follow a power-law). To create the high-mass end of the synthetic CMF ($M > M_{\rm{min}}$), we generated $n_{\rm{high}}$ synthetic cores more massive than $M_{\rm{min}}$ that were distributed according to the power-law fit to the observed CMF. This ensured that the high-mass end of the synthetic sample is guaranteed to follow a power-law (although apparent deviations can arise due to random sampling of a finite population). We then combined the low-mass and high-mass samples into a full synthetic CMF covering the entire mass range of the observed CMF.

We note that to create a synthetic CMF for the true CMF, which by definition does not possess any information about the distribution of cores within each mass bin, we generated the corresponding number of masses uniformly distributed across each mass bin.  We then performed the sampling procedure described above and re-binned the resulting synthetic sample according to the binning scheme originally used to create the binned CMFs described in \S\ref{S:cmfcon}. 

Once a given synthetic sample was created, we fitted a power-law to it using the fiducial MLE method, including fitting an optimal $M_{\rm{min}}$.  By creating a sample which is guaranteed to follow a power-law above a certain mass, but fitting a power-law without knowledge of what that mass was, we replicate the conditions inherent to our fitting process for the real CMFs.  Due to the random fluctuations in the shapes of the synthetic CMFs this allows their fits to yield $M_{\rm{min}}$s that are greater or less than the $M_{\rm{min}}$ actually used to create that CMF.  This preserves the integrity of the goodness of fit calculation.  We then repeated this procedure $N=3000$ times, i.e., with 3000 different synthetic CMFs for each $M_{\rm min}$, recording the value of $\hat{D}$ or $\hat{A}^2$ resulting from the power-law fit to each synthetic CMF.

We calculated the fraction of synthetic values of $\hat{D}$ or $\hat{A}^2$ that were larger than the original $D$ or $A^2$ fit to the real CMF and we estimate a $p$-value via $p = N(\hat{D} \geq D)/N$ and $p = N(\hat{A}^2 \geq A^2)/N$.  Since these statistics are meant to be minimized, this represents the percent of synthetic fits which, although known to actually follow a power-law above $M_{\rm min}$, resulted in a ``worse’’ fit, i.e., due to random sampling effects.  This percentage is equivalent to the $p$-value for the fit to the real CMF; if many synthetic CMFs randomly appear to be a worse fit, it indicates that the real CMF is likely to be well-described by said power-law.  Alternately, if very few of the synthetic CMFs (e.g., $<10\%$, i.e., $p < .10$) result in fits which are worse than the fit to the real CMF, this indicates that the real CMF is unlikely to be described by that distribution.  

For the raw and flux-corrected CMFs, we performed this bootstrapping procedure for 100 values of $M_{\rm{min}}$ evenly distributed between 0.1 and 10 $M_{\odot}$.  For the true CMF, we performed this procedure for 11 mass bins between 0.1 and 10 $M_{\odot}$.
We note that the power of this test decreases as $M_{\rm{min}}$ is raised and $n_{\rm{high}}$ decreases, which effectively increases uncertainties on individual $p$-values at higher $M_{\rm{min}}$'s.

Figures \ref{fig:d_mmins}c and \ref{fig:d_mmins}d show the $p$-values of power-law fits to the CMFs as a function of $M_{\rm{min}}$ for the $D$ and $A^2$ statistics, respectively.  
The black and blue curves, corresponding to the raw and flux-corrected CMFs, respectively, were smoothed using LOWESS regression (Locally Weighted Scatter-plot Smoothing) \citep{cleveland_robust_1979} with a smoothing parameter of 0.05 determined by cross-validation. This method fits polynomials to many small subsets of the data through WLS regression to reduce noise in the data and make visual assessment of trends easier. 

We find that a power-law becomes a consistently plausible fit ($p \geq .10$) to the raw CMF above $\sim 3 \ M_{\odot}$ as assessed by the KS statistic and above $\sim 3.6 \ M_{\odot}$ as measured by the AD statistic, and is not adequate at describing the CMF below these masses.  A power-law becomes appropriate for the flux-corrected CMF above $\sim 6 \ M_{\odot}$ as indicated by both the KS statistic and the AD statistic.  We find no significant evidence that power-law behavior holds below this range.  A power-law becomes an adequate descriptor of the histogram of the true CMF above $\sim 5 \: M_{\odot}$ as assessed by the KS statistic and also above $\sim 5 \: M_{\odot}$ as measured by the AD statistic. 

Considering the various results derived from the KS and AD based methods, we report final ``best'' fits that begin at the lowest significant ($p \geq .10$) $M_{\rm min}$ value as assessed by $A^2$.  We prefer the AD statistic due to the known issue of the KS statistic being insensitive to differences in the tails of compared distributions.  The single power law best fit to the raw CMF is then 
$\alpha=0.95 \pm 0.09$ with $M_{\rm{min}}$ $\sim 3.6 \ M_{\odot}$  ($p=.15$).  The best fit to the flux-corrected CMF is $\alpha=1.04 \pm 0.11$ above $M_{\rm{min}}$ $\sim 6.0\ M_{\odot}$ ($p=.11$).  The best fit to the true CMF is $\alpha = 0.94 \pm 0.08$ above the bin with a lower limit of $M_{\rm{min}}$ $\sim 5.0 \ M_{\odot}$ ($p=.12$).  These results are presented in Table~\ref{tab:plawparams}.  

\subsubsection{PPL fits and evidence for a break in the CMF}\label{S:break}

\begin{figure}
    \centering
    \includegraphics[width=0.45\textwidth]{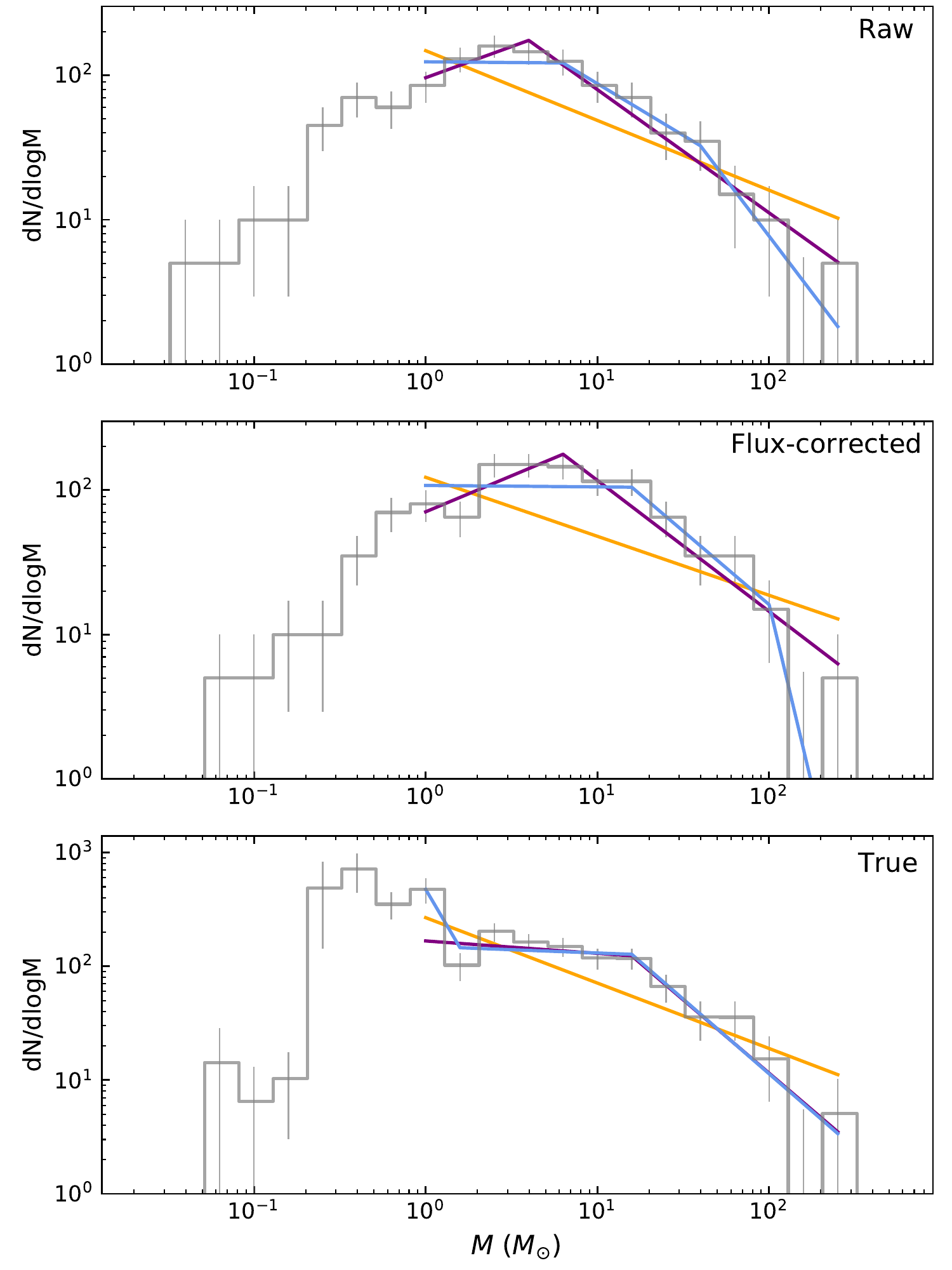}
    \caption{\textit{(a) Top:} Multi-component piecewise power-law (PPL) fits to the raw CMF, shown by the grey histogram, with Poisson errors indicated. The orange, purple and blue lines are the best single, double and triple PPL fits, respectively. All cases shown here were fit from the bin a $1\:M_\odot$ to a maximum bin centered at about $250\:M_\odot$. \textit{(b) Middle:} As (a), but for the flux-corrected CMF. \textit{(c) Bottom:} As (a), but for the true CMF.}
    \label{fig:trucmf_broke}
\end{figure}

The results of the last section already indicate that a single power law description of the true CMF is only valid at the high-mass end above about $5\:M_\odot$. Here we compare the goodness of fit of single and multi-component continuous power laws, i.e., piecewise power laws (PPLs), to the CMF to further investigate if a break in the single power law is present and what are the best PPL descriptions. PPL descriptions of the CMF and IMF have been made previously in a number of studies \citep[e.g.,][]{johnstone_large-area_2000,kroupa_variation_2001,reid_highmass_2006}.

For the mass ranges defined by minimum mass bins $M_{\rm{min}}$ centered from 1 and 16 $M_{\odot}$ we fit a single power-law, a double PPL, and a triple PPL by performing non-linear WLS regression. The double PPL can be expressed as:
\begin{equation}
    p(M) = 
    \begin{cases}
    A M^{-\alpha_{1}} & M < M_{1} \\
    A (M_{1}^{-\alpha_{1}+\alpha_{2}}) M^{-\alpha_{2}} & \text{otherwise}
    \end{cases}
\end{equation}
and the triple PPL as:
\begin{equation}
    p(M) = 
    \begin{cases}
    A M^{-\alpha_{1}} & M < M_{1} \\
    A (M_{1}^{-\alpha_{1}+\alpha_{2}}) M^{-\alpha_{2}} & M_{1} \leq M \leq M_{2} \\
    A (M_{1}^{-\alpha_{1}+\alpha_{2}}) (M_{2}^{-\alpha_{2}+\alpha_{3}}) M^{-\alpha_{3}} & M_{2} < M
    \end{cases}
\end{equation}
The different PPL fits resulting from this method for $M_{\rm min} = 0.79\: M_{\odot}$ are shown in Figure \ref{fig:trucmf_broke}.  Note, all fits include the entirety of the high-mass end of the CMF up to the highest mass core $M=252\:M_{\odot}$, with the empty bin at 125 $M_{\odot}$ being assigned a count as in the original WLS fitting in \S\ref{S:fitting}.  

\begin{figure}
    \centering
    \includegraphics[width=0.45\textwidth]{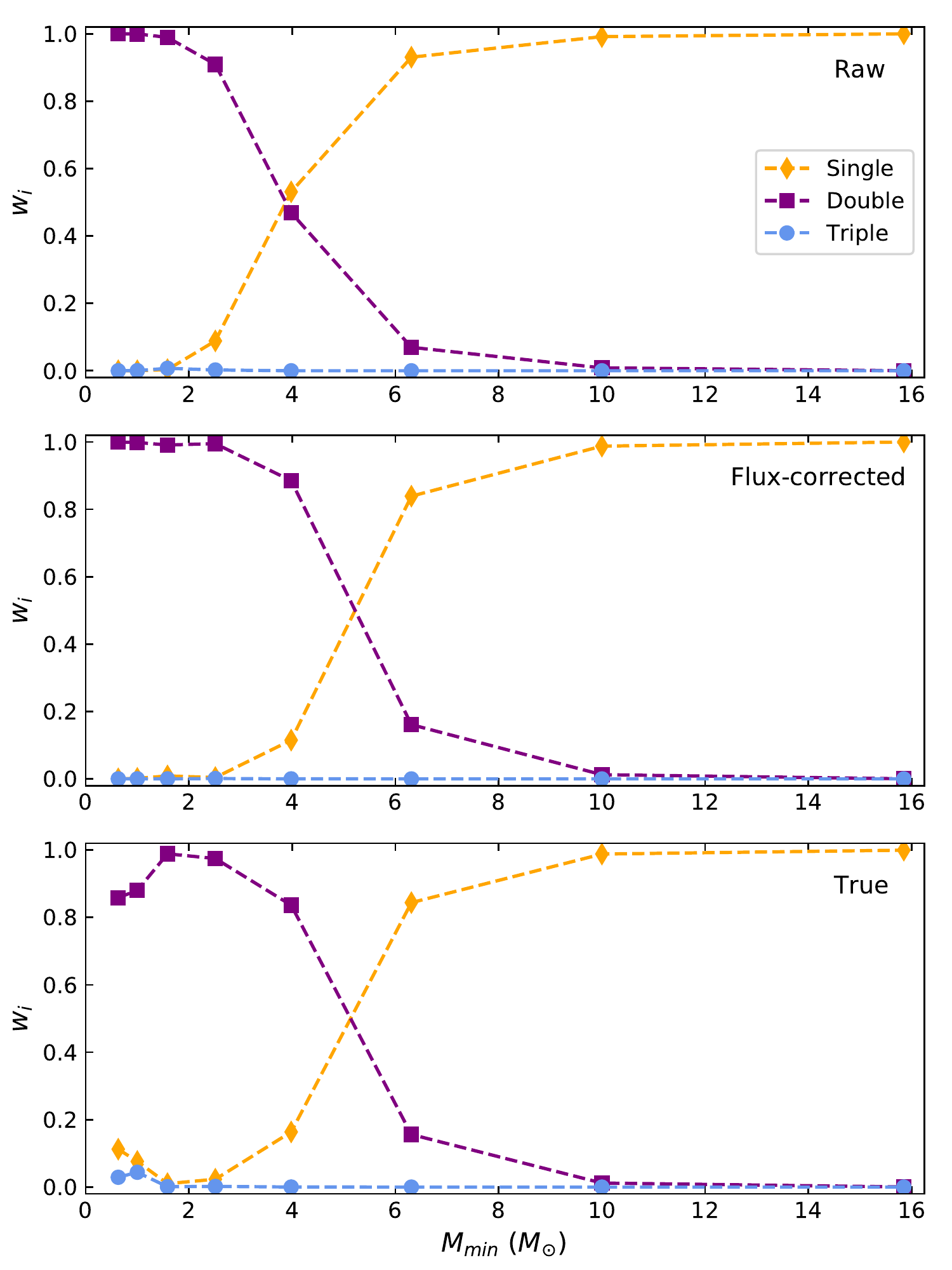}
    \caption{\textit{(a) Top:} The Akaike weight \textit{$w_{i}$} of each fit to the raw CMF versus the $M_{\rm{min}}$\ from which it was derived. The single, double and triple PPL fit results are shown in orange, purple and blue, respectively. \textit{(b) Middle:} As (a), but for the flux-corrected CMF. \textit{(c) Bottom:} As (a), but for the true CMF.}
    \label{fig:prob_aic}
\end{figure}

The Akaike Information Criterion (AIC) \citep{akaike1973information} can be used to choose a best fit model for a given data set by measuring the ``distance'' between the observed and modeled distributions. For WLS estimation, this takes the form:
\begin{eqnarray}
    \rm{AIC}_{WLS} & = & N\ {\rm ln} \left( \frac{ \sum_{j=1}^{N} w_{j}^{-2} (y_{i} - f(t_{j},q_{\rm WLS}))^{2} } {N} \right)\nonumber\\ 
    & & + 2(\kappa_{q}+1),
\end{eqnarray}
where $N$ is the number of bins being fit and $\kappa_{q}$ is the number of parameters being fit \citep{banks_aic_2017}.  For a single, double, and triple PPL $\kappa_{q}=2$, $4$ and $6$, respectively.  $N$ ranges between 7 and 13 depending on the $M_{\rm{min}}$ selected. However, for sample sizes $N$ that are small in comparison to $\kappa_{q}$, as occurs here, the AIC is known to perform poorly. The corrected Akaike Information Criterion AIC$_{c}$ \citep{doi:10.1080/03610927808827599} accounts for this effect by adding an additional penalty term to the AIC formula:
\begin{equation}
     {\rm AIC}_{\rm WLS,c} = {\rm AIC}_{\rm WLS} + \frac{2(\kappa_{q} + 1)(\kappa_{q} + 2)}{N-\kappa_{q}}.
\end{equation}
We calculate the single, double and triple PPL fit AIC$_{\rm WLS,c}$ values as a function of \bmin for each CMF.  This allowed for analysis of the relative likelihood of these forms as a function of $M_{\rm{min}}$.  We define the AIC differences $\Delta_{i}$(AIC) \citep{wagenmakers_aic_2004} as

\begin{equation}
    \Delta_{i}(\rm{AIC}) = \rm{AIC}_{i} - \rm{AIC}_{min},
\end{equation}
where $\rm{AIC}_{min}$ is the minimum AIC$_{\rm WLS,c}$ of the three fits.  The likelihood of a model is proportional to exp$(-\frac{1}{2} \Delta_{i})$, which allows for the calculation of Akaike weights $w_{i}$(AIC),

\begin{equation}
    w_{i}({\rm AIC}) = \frac{{\rm exp}\left(-\frac{1}{2} \Delta_{i} \right)}{\sum_{k=1}^{K} {\rm exp}\left(-\frac{1}{2} \Delta_{k} \right)},
\end{equation}
which are the normalized relative likelihoods of that model and where $K$ is the number of candidate models.  This is equivalent to the probability that each model is the ``best'' model.  The  $w_{i}$(AIC) values for all model types fit to the three CMFs as a function of $M_{\rm{min}}$ are presented in Figure \ref{fig:prob_aic}. 

We find that for low values of $M_{\rm min}\sim 1\:M_\odot$ all CMFs prefer a double PPL. However, they eventually transition to strongly favor a single power-law as $M_{\rm{min}}$ is raised.  This turnover occurs at about $4\:M_{\odot}$ for the raw CMF, $5\:M_{\odot}$ for the flux-corrected CMF, and also $5\:M_{\odot}$ for the true CMF.  A triple PPL is never indicated as being more appropriate than the other forms.

When fitting over the ``standard'' range of the CMF with $M_{\rm{min}}$ $=0.79\: M_{\odot}$, the best double PPL fit to the raw CMF is where $M_{1} = 4 \: M_{\odot}$, $\alpha_1 = -0.43 \pm 0.13$, and $\alpha_2 = 0.85 \pm 0.07$.  
The flux-corrected CMF fitting results are similar.
For the true CMF the best fit results are $M_{1} = 15 \: M_{\odot}$, $\alpha_1 = 0.11 \pm 0.12$, and $\alpha_2 = 1.29 \pm 0.26$. Note, the break mass bin $M_1$ for the raw and flux-corrected CMFs are identical to the bins selected by fitting a single power-law to the binned CMF through performing an MLE and minimizing $A^2$.  
This reinforces that a break in the CMF exists in this mass range. The break mass for the true CMF is higher due to the effects of number corrections inflating the number of cores in the lower mass bins near $1 \: M_{\odot}$.

We conclude that this consistent transition from a double to single power-law as $M_{\rm{min}}$ increases 
indicates the presence of a break in the CMF between about 4 and $15\:M_{\odot}$, 
depending on the overall range of the CMF that is fit and being moderately sensitive to whether number corrections are applied to the CMFs. These results are consistent with those derived from the KS and AD based methods, i.e., which also support a break in power-law behavior of the CMF around 4 to $6\:M_{\odot}$.

\subsubsection{Comparison to log-normal distributions}\label{S:comp_dists}

\begin{figure*}
    \centering
    \includegraphics[width=\textwidth]{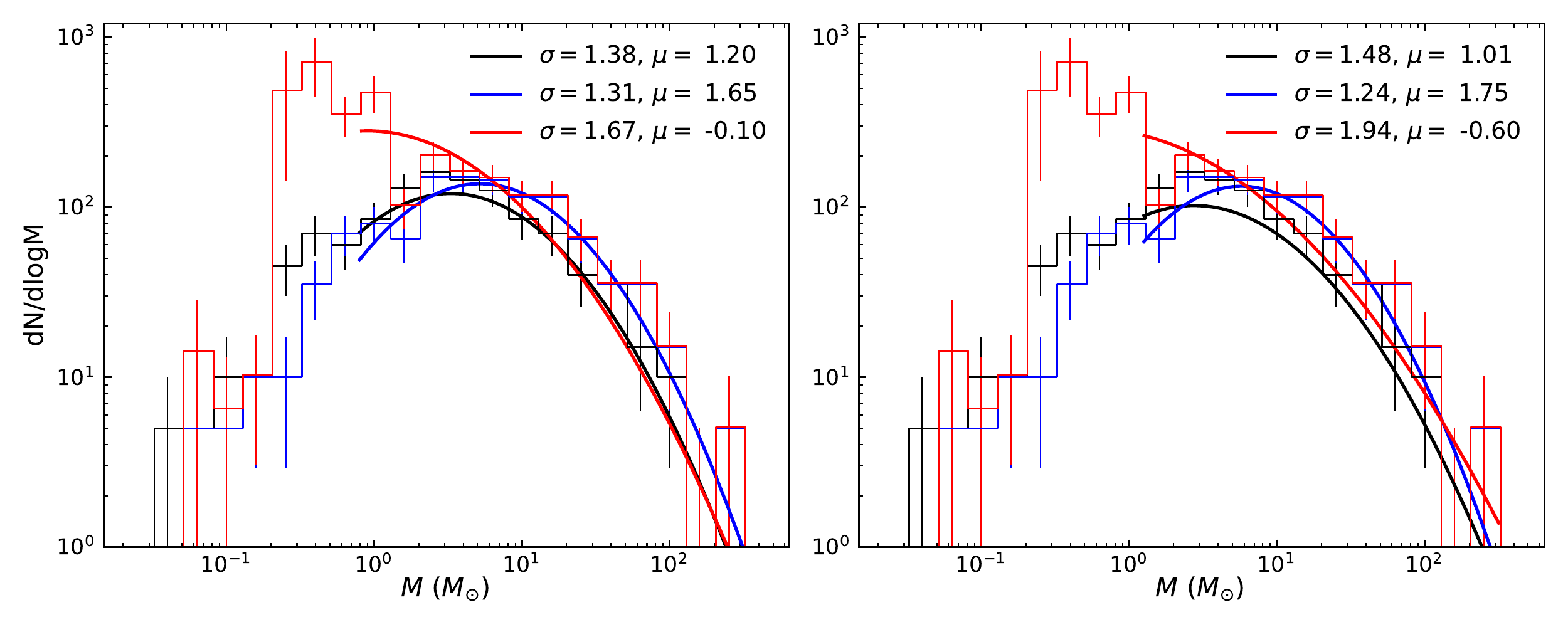}
    \caption{Best log-normal fits to the CMF.  \textit{(a) Left:} The black histogram shows the raw CMF, the blue histogram shows the flux-corrected CMF, and the red histogram shows the true CMF. The solid black, blue and red lines show the best log-normal fits derived through MLE above $M_{\rm{min}}=0.79$ $M_{\odot}$.  \textit{(b) Right:} As (a), but for $M_{\rm{min}}=1.25$ $M_{\odot}$}
    \label{fig:lognorm_cmf}
\end{figure*}

We next consider log-normal fits to the CMF. Such log-normal descriptions of the IMF and/or CMF have been discussed by e.g., \citet{larson_simple_1973,zinnecker_star_1984,adams_theory_1996}.
The log-normal probability density function takes the form
\begin{equation}
    p(M) = \frac{1}{\sigma M \sqrt{2\pi} } {\rm exp} \left( - \frac{({\rm ln} \ M-\mu)^2}{2\sigma^2} \right),
\end{equation}
where $\mu = {\rm ln} \bar{M}$ with $\bar{M}$ being the mean mass and $\sigma^2$ is the variance in ${\rm ln}\:M$, i.e., $\sigma$ describes the width of the distribution in ${\rm ln}\:M$.  

\citet{chabrier_galactic_2003} found $\sigma \sim 0.7$ for the IMF.  This was later revised to $\sigma \sim 0.55$ \citep{chabrier_2005}. Subsequently, \citet{bochanski_luminosity_2010} derived $\sigma \sim 0.3$ for the IMF. Previous studies of the CMF have derived values of $\sigma$ between 0.6 and 2 \citep[see e.g.,][]{reid_highmass_2006,swift_discerning_2010,olmi_shape_2013,konyves_census_2015}.  

Fitting only for $M>0.79\:M_\odot$, we derive the log-normal fits shown in Figure~\ref{fig:lognorm_cmf}a. The results of fitting using the next higher mass bin as the minimum are shown in Figure~\ref{fig:lognorm_cmf}b. We find that the global log-normal fit is quite sensitive to the number corrections applied to transform the flux-corrected CMF into the ``true'' CMF. These corrections become relatively large for the bin centered at $1\:M_\odot$ and so the fit parameters are also quite sensitive to whether this bin is used in the fitting. For example, the mean core mass is $\bar{M}=0.90\:M_\odot$ when the fit is carried out down to this bin and $\bar{M}=0.55\:M_\odot$ (and with a somewhat broader log-normal) when going down to the next higher mass bin.

We explored the goodness of fit of log-normal distributions and single power-laws by calculating the ratio $R$ of their log-likelihoods \citep{vuong_likelihood_1989,clauset_power-law_2009}:
\begin{equation}
R = \sum_{i=1}^{n} [{\rm ln} \: p_{\rm pl}(M_{i}) - {\rm ln} \:  p_{\rm ln}(M_{i})].
\end{equation}
The significance levels of this test were calculated through the {\tt{powerlaw}} Python package \citep{alstott_powerlaw_2014}.  We report values of $R$ normalized by its standard deviation, $R/(\sigma_R \sqrt{n})$.  To assess the binned CMFs, we generated the corresponding number of random masses uniformly distributed in each mass bin and calculated the median $R$ over 2000 iterations of the same process.  In our implementation of this test, positive values of $R$ indicate that a power-law is preferred, while negative values indicate that a log-normal distribution is preferred.  $p$-values larger than $p=.10$ indicate that there is no significant difference between the goodness of fit of the two distributions.

For $M_{\rm min}= 0.79\:M_{\odot}$, a log-normal distribution is a significantly better fit to the raw and flux-corrected CMFs than a single power-law ($R=-4.6$, $p<.001$ for the raw CMF; $R=-5.7$, $p<.001$ for the flux-corrected CMF). However, there was no significant evidence that a log-normal was a better descriptor of the true CMF than a single power-law ($R=-1.2$, $p=.25$). When fitting with the next bin in mass as the minimum, i.e., $M_{\rm min}= 1.26\:M_{\odot}$, a log-normal distribution remained more appropriate than a single power-law for the raw and flux-corrected CMFs ($R=-3.2$, $p=.002$ for the raw CMF; $R=-5.4$, $p<.001$ for the flux-corrected CMF). However, there continued to be no significant evidence that a log-normal fit was more appropriate for the true CMF than a single power-law ($R=-1.03$, $p=.30$). 

We also compared a log-normal fit to PPLs using the methods of \S\ref{S:break}.  We calculated the AIC$_c$ values and Akaike weights $w_{i}$(AIC) for a log-normal fit and for single, double, and triple power-law fits to the CMF.  With $M_{\rm min}= 0.79\:M_{\odot}$, the raw and flux-corrected CMFs were both best described by a log-normal fit, with $w_{i}$(AIC)$_{\rm{ln}}=0.97$ and $w_{i}$(AIC)$_{\rm{ln}}=0.99$, respectively. However, for the true CMF a log-normal fit was not the best description ($w_{i}$(AIC)$_{\rm{ln}}=0.03$) and a double PPL was preferred ($w_{i}$(AIC)$_{\rm{double}} = 0.85$). Above $M_{\rm{min}}=1.26 \: M_{\odot}$,  
a PPL was still preferred for the true CMF ($w_{i}$(AIC)$_{\rm{double}} = 0.97$).  

In summary, these results indicate that a single log-normal is not a better fit of the true CMF compared to a double PPL, at least over the range of masses explored in this study, down to about $1\:M_\odot$. We do note, however, that single log-normals are relatively good fits to both the raw and flux-corrected CMFs, so this result for the true CMF is dependent on the accuracy of our estimated incompleteness corrections. Other models, such as hybrid models with a log-normal at low masses plus a high-mass power law tail are likely to be possible good descriptions of the true CMF. However, to advance further a CMF measurement with larger numbers of cores (to reduce Poisson errors) and probing to lower masses (to more accurately define the expected peak) are needed.

\subsubsection{Effects of Dendrogram Parameters, Binning Scheme, and Clump Distance}\label{S:dendro}

We varied the parameters used to identify cores with the dendrogram to test the effects on our results.  Our fiducial method is with a minimum intensity $F_{\rm{min}}=4\sigma$, a minimum increase in intensity $\delta = 1 \sigma$, and a minimum area $S_{\rm{min}}=0.5$ of the size of the synthesized beam.  We kept $\delta$ constant and tested combinations of $F_{\rm{min}}=$ 3, 4, or 5 $\sigma$ and $S_{\rm{min}}=0.5$ beam or 1 beam.  We found a minimum of 128 cores with $F_{\rm{min}}=5\sigma$ and $S_{\rm{min}}=1$ beam, and a maximum of 293 cores with $F_{\rm{min}}=3\sigma$ and $S_{\rm{min}}=0.5$ beam.  For the true CMF, $M_{\rm{min}}$ typically ranged between 3 and $8\:M_{\odot}$ with $\alpha$'s between 0.94 and 1.10.  This shows that our results using the fiducial method are not very sensitive to the choice of dendrogram parameters.

Fitting to histograms of the CMF can create a bias in the derived values of $\alpha$ through the information loss created by binning.  The fiducial binning method involves five bins per dex.  We varied the number of bins per dex and fit a power-law to the resultant histogram using the fiducial binned MLE method (MLE-B). As expected, we found that the $\alpha$ derived for each histogram converged on the values derived from fitting to the unbinned sample of cores as the number of bins per dex increased.
To be consistent with the methods of Papers I and II we report values derived from fitting to histograms with five bins per dex, but note the potential bias in these binned MLE based results. The results shown in Table 2 comparing MLE and MLE-B results indicate that these biases are present at a level corresponding to $\Delta\alpha\simeq 0.1$, which is about the uncertainty level caused by Poisson sampling uncertainties of the CMFs.

\begin{figure*}
    \centering
    \includegraphics[width=\textwidth]{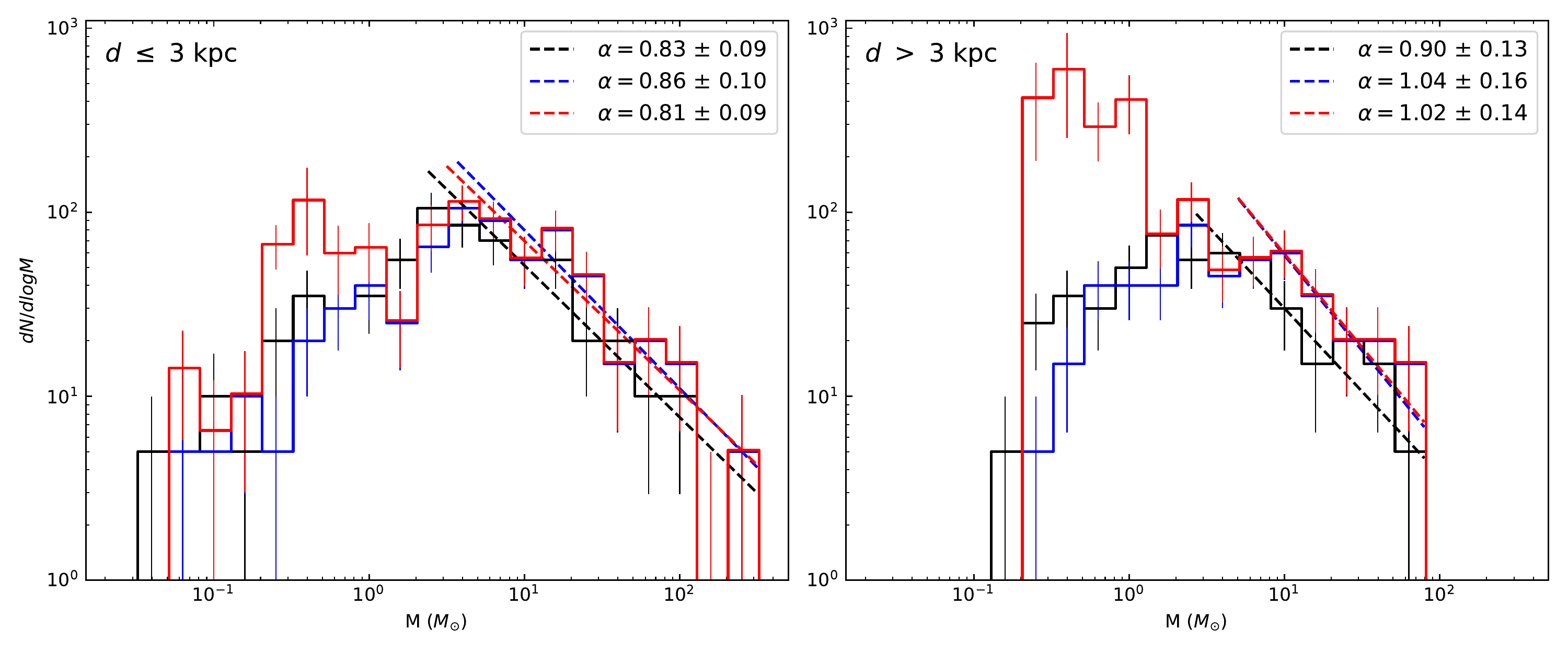}
    \caption{\textit{(a) Left:} The combined CMF of the 15 clumps within 3 kpc. The black histogram shows the original, ``raw'' CMF.  The blue histogram shows the CMF after flux correction and the red histogram shows the final, ``true'' CMF, i.e., after applying number recovery fraction correction to the flux-corrected CMF. The dashed black, blue and red lines show the best power law fits derived through the fiducial MLE fitting process, beginning at  $M_{\rm{min}}=2.4, 3.7,$ and $3.1\:M_{\odot}$, respectively. \textit{(b) Right:} Same as the left panel, but for the 13 clumps farther than 3 kpc.  The dashed lines correspond to the best power-law fits derived through the fiducial MLE fitting process (see text), beginning at $M_{\rm{min}}=2.7, 5.0$ and $5.0\:M_{\odot}$, respectively.}
    \label{fig:cmfs_distance}
\end{figure*}

We also examined the variation in the CMF as a function of clump distance.  Clumps distances ranged between 0.79 kpc and 3.5 kpc.  Figure \ref{fig:cmfs_distance} compares the CMFs of the 15 clumps (containing 127 cores) within 3 kpc and the 13 clumps (containing 95 cores) farther than 3 kpc.  Unsurprisingly, we are able to probe to lower mass cores in the near sample. The far sample contains a handful of cores that are very close to the noise limit, subject to very large number correction factors $\gtrsim20$, and that are mainly responsible for the dramatically boosted ``true'' CMF at $\lesssim 1\:M_\odot$ in the far sample. This gives us further reason to be cautious about the reliability of the derived CMF in this regime. Conversely, the shape of the CMF of the near sample, which is subject to smaller number corrections in the $\sim 1\:M_\odot$ regime, strengthens our confidence in the result that a break is seen in the CMF at masses $\gtrsim 5\:M_\odot$.

We performed KS tests comparing the raw and flux-corrected CMFs of the near and far clumps, and found no significant evidence that the samples are drawn from different populations ($p=.09$ and $p=.07$, respectively).  We did find significant differences between the true CMFs ($p<.001$), which is likely the result of the increased effect of number corrections in the far sample.

We fit single power-laws to the near and far CMFs using the fiducial MLE fitting process and found good agreement with the fits to the complete CMFs. The fitted values of $\alpha$ are largely within the estimated errors of $\alpha$ in the original fits. The values of $M_{\rm{min}}$ ranged between $\sim 2 - 4 \ M_{\odot}$ for the near sample and $\sim 3 - 5 \ M_{\odot}$ for the far sample. Thus, note that both samples independently show evidence for a break in the CMF via this method.

\subsection{Spatial Distribution of Cores}\label{S:spatial}

\begin{figure}
    \centering
    \includegraphics[width=0.48\textwidth]{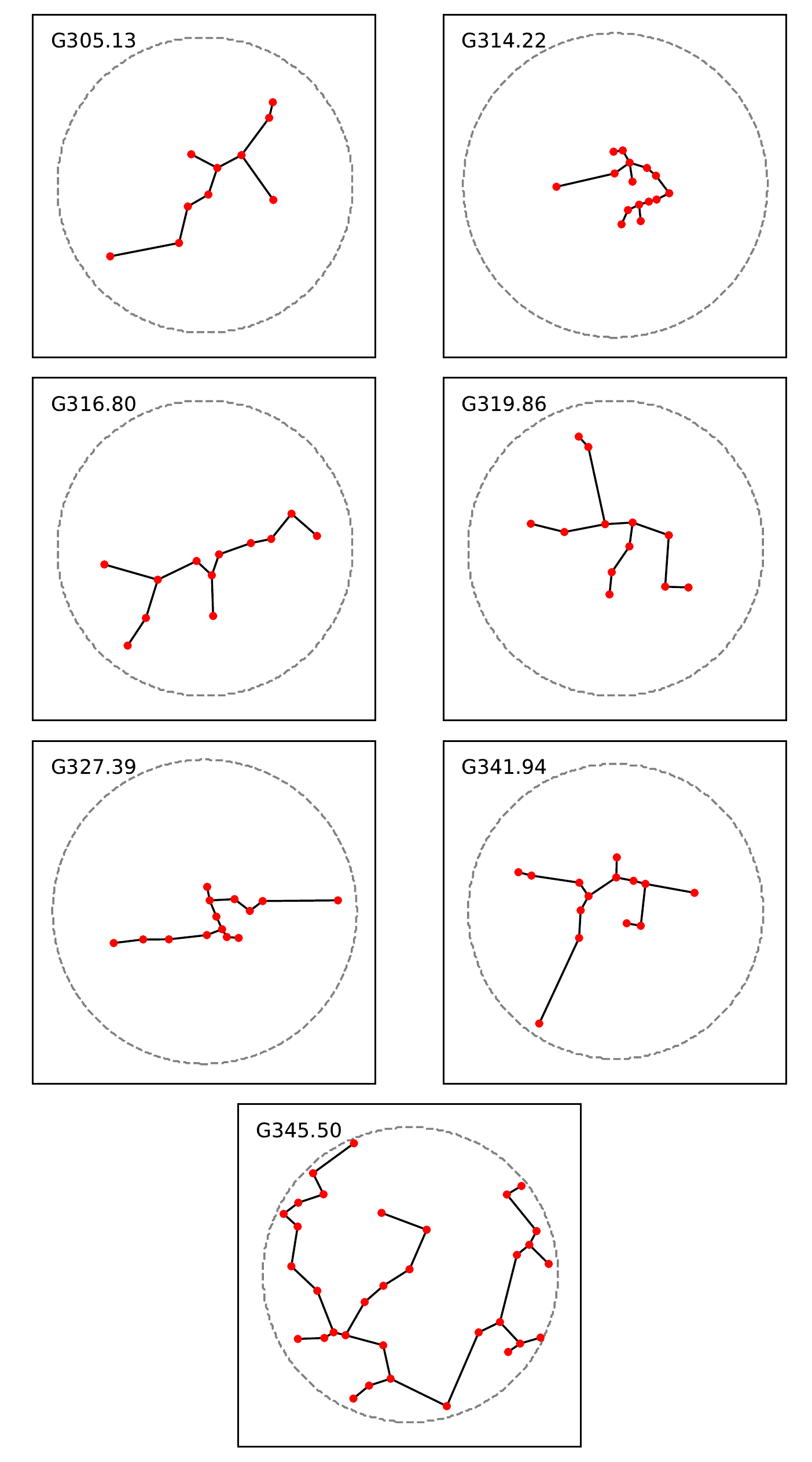}
    \caption{Minimum spanning trees (MSTs) calculated for the seven clumps each containing at least ten cores.  The extent of the ALMA primary beam FOV is marked by the dashed circle.  The location of each core is represented by the red dots, and the edges of the MST between cores are shown by the black lines.}
    \label{fig:msts}
\end{figure}

For those clumps with at least 10 cores, we analyzed the clustering of cores within clumps by calculating minimum spanning trees (MST), i.e., the set of edges connecting a set of points without closed loops where the sum of the edge lengths is a minimum \citep{kruskal_shortest_1956,prim_shortest_1957}.
The resulting MSTs, which were generated using the Python package $\tt{networkx}$ \citep{SciPyProceedings_11}, are shown in Figure\,\ref{fig:msts}. For each clump, we calculated the dimensionless $Q$ parameter \citep{cartwright_statistical_2004},
\begin{equation}
Q = \frac{\bar{m}}{\bar{s}},
\end{equation}
where $\bar{m}$ is the normalized mean edge length of the MST and $\bar{s}$ is the normalized mean edge length between all cores. With some variation with the density profile assumed, large $Q$ values ($Q>0.8$) indicate that structures are centrally concentrated, while small $Q$ values ($Q<0.8$) indicate that clusters have fractal substructures. The use of MSTs and the $Q$ parameter in the context of dense cores has been extensively studied in previous works \citep[see e.g.,][]{schmeja_evolving_2006,wu_gmc_2017,sanhueza_alma_2019,dib_star_2019,lu_alma_2020}.
\begin{deluxetable}{ccccc}
\tablewidth{0.45\textwidth}
\tablecaption{Core Clustering Properties\label{qparams}} 
\tablehead{ \colhead{ID} &\colhead{$N_{c}$}& \colhead{$\bar{m}$} & 
\colhead{$\bar{s}$} & \colhead{$Q$}
}             
\startdata  
G305.13 &     10 &  0.46 &  0.69 &  0.67 \\\ G314.22 &     15 &  0.48 &  0.59 &  0.82 \\\ G316.80 &     12 &  0.54 &  0.85 &  0.64 \\\ G319.86 &     12 &  0.61 &  0.84 &  0.72 \\\ G327.39 &     14 &  0.38 &  0.55 &   0.70 \\\ G341.94 &     14 &  0.45 &  0.59 &  0.77 \\\ G345.50 &     33 &  0.67 &  0.98 &  0.69
\enddata      
\end{deluxetable}

Seven out of 28 clumps in our sample contain at least ten cores. 
We found little evidence for centrally condensed structures in these regions, with values of $Q$ ranging from 0.64 to 0.82, as summarized in Table\, \ref{qparams}. However, we note that the relatively small sample sizes of cores within each clump severely limit the conclusions that can be derived from these results.  \citet{cartwright_statistical_2004} identified a reduction of the accuracy of this method below $N=200$ members, while \citet{gouliermis_clustered_2012} found through the use of simulated clusters that clusters containing $N<30$ members experience especially pronounced degradation in the accuracy of $Q$ estimations.

Additionally, the small number of cores in each clump prevented the calculation of the frequently used ``mass segregation ratio'' $\Lambda_{MSR}$ and other typical methods to investigate this effect on a clump-by-clump basis.  We compared all core masses to the radial separation of each core from the center of its image and observed a slight, although non-significant trend of higher mass cores being located nearer to the center of images ($r=-0.12$).

\section{Comparison of CMFs in Different Environments and Implications for Massive Star Formation}\label{S:discussion}

\begin{figure}
    \centering
    \includegraphics[width=0.45\textwidth]{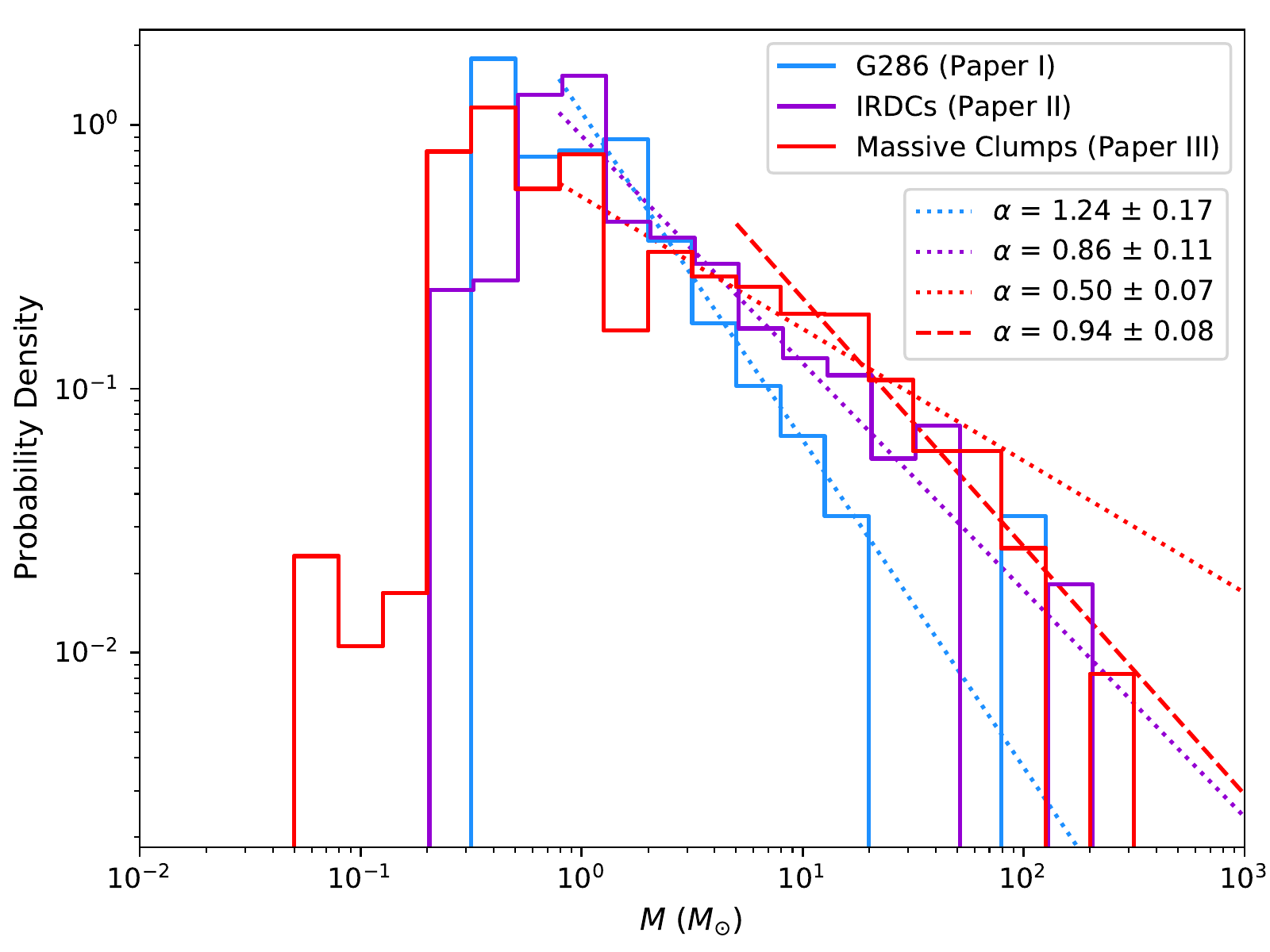}
    \caption{The blue histogram is the true CMF derived from the G286 protocluster in Paper I.  The purple histogram is the combined true CMF derived from the seven IRDCs in Paper II.  The red histogram is the combined true CMF from this sample of massive clumps.  The three histograms are each normalized by the total number of cores within each CMF.  The dotted lines show the best WLS power-law fits for each CMF with $M_{\rm{min}}$ $=0.79$ $M_{\odot}$, with colors corresponding to each histogram. The dashed red line shows the best fit to the sample of massive clumps with $M_{\rm{min}}$ $=5$ $M_{\odot}$.}
    \label{fig:Comp_cmfs}
\end{figure}

We have derived CMFs in different types of protoclusters using the same methods. This now enables direct comparisons to be made between them.
Figure \ref{fig:Comp_cmfs} compares the 
``true'' CMFs that we have derived for the massive protocluster sample to those derived in Paper I for the G286 protocluster and Paper II for the IRDC clumps. 

We note that using the method of maximum likelihood, we derive $\alpha = 1.45 \pm 0.15$ for G286 when $M \geq 0.79 \ M_{\odot}$ and $\alpha = 1.03 \pm 0.07$ for the sample of IRDCs for $M \geq \ 0.79 \ M_{\odot}$. These values are modestly steeper than the WLS reported results from Papers I and II. For the CMF found here in Paper III, we derive a significantly shallower $\alpha = 0.60 \pm 0.05$ for $M \geq 0.79 \ M_{\odot}$. However, as we have discussed, this single power law fit is not an good description of the distribution and a double PPL is favored, i.e., with a break around $\sim 3$ to 10~$M_\odot$.

We performed Kolmogorov-Smirnov tests comparing CMFs from our previous studies in G286 and IRDCs to the CMF derived here in the ALMAGAL-observed massive protoclusters. We only include cores with $M \geq 0.79$ $M_{\odot}$, i.e., more massive than the lower bound of the bin centered at $1\:M_{\odot}$, in these tests in order to mitigate the effects of the large uncertainties on the true CMFs below this mass range due to large number correction factors. From this analysis we find significant differences between the populations. The true CMF is significantly different than the true CMFs of Paper I ($p < .001$) and Paper II ($p < .001$).

The nature of the difference is that we find that the CMF from our sample of ALMAGAL-observed massive protoclusters has a significantly larger population of high-mass cores than the CMFs derived from the G286 protocluster and the IRDC clumps. G286 is known to be in a relatively evolved (although still gas dominated) state of formation, while the IRDC clumps are likely to be at an earlier stage. On the other hand, the massive protoclusters have been selected based on their having relatively strong sub-mm emission, without regard to any evolutionary state, although they tend to be actively forming stars, rather than being prestellar clumps (see Table~\ref{tab:sources}). Other properties of the massive protoclusters, i.e., $R_{\rm cl}$, $M_{\rm cl}$ and $\Sigma_{\rm cl}$, are also listed in Table~\ref{tab:sources}. In comparison to G286 and the IRDC clumps, the main difference is the typically much higher value of mass surface density, $\Sigma_{\rm cl}$, of the massive protocluster sample: these sources generally have $\Sigma_{\rm cl}>1\:{\rm g\:cm}^{-2}$ (although this is dependent on the scale of the region considered), while the IRDC clumps usually have $\Sigma_{\rm cl}\lesssim 0.4\:{\rm g\:cm}^{-2}$ \citep{butler_mid-infrared_2012}. On scales within a radius of 1.54~pc, relevant to where its core population is measured, G286 is estimated to have $\Sigma_{\rm cl}\sim 0.1\:{\rm g\:cm}^{-2}$ \citep{cheng_gas_2020}.  Thus, we conclude that this difference in environmental conditions is the factor that is most likely to be responsible for the variation in the observed CMFs.

Next, one may speculate on the reason why high $\Sigma_{\rm cl}$ leads to a more top-heavy CMF. At fixed temperature, the critical Bonnor-Ebert mass is given by $M_{\rm BE} = 0.0504 (T/20 \ \rm{K})^2 (\Sigma_{\rm cl}/1\:{\rm g\:cm}^{-2})^{-1}$ \citep[see their eq.~68]{mckee_formation_2003}, i.e., scaling as $\Sigma_{\rm cl}^{-1}$, i.e., becoming much smaller in high pressure regions (in a self-gravitating clump the ambient pressure $\propto\Sigma_{\rm cl}^2$). In any case, for realistic temperatures $\lesssim100\:$K and the relevant range of $\Sigma_{\rm cl} \gtrsim 1$ g cm$^{-2}$ the Bonner-Ebert mass is $\lesssim 1\:M_\odot$, and so is not related to the differences we see that apply to the shape of the CMF in the region from $\sim1$ to $100\:M_\odot$. Differences in the fragmentation scale in this mass regime would likely require there to be differences in the strength of magnetic fields, which do appear likely to be influencing fragmentation in IRDCs \citep[see e.g.,][]{butler_mid-infrared_2012,beuther_magnetic_2018}.

Alternatively, the core properties may be evolving due to enhanced accretion from the ambient clump gas and/or mergers between cores. These processes are expected to become more important in high $\Sigma_{\rm cl}$ environments. For example, the Turbulent Core Model of \citet{mckee_formation_2003} predicts that the rate of accretion of a core from the ambient clump is $\propto \Sigma_{\rm cl}^{3/4}$ and at typical values of $\Sigma_{\rm cl}\sim 1\:{\rm g\:cm}^{-2}$ is already occurring at a rate that would double the mass of a core within about one core free-fall time. Thus, this mechanism of core growth from the clump, especially in the build up of prestellar cores that are relatively long-lived, could be important in shaping the CMF that we observe, which in these mm-continuum selected samples is likely dominated by protostellar cores.

We emphasize here that scenarios in which prestellar core properties evolve significantly due to accretion from their surroundings or mergers with other cores are not to be confused with Competitive Accretion \citep{bonnell_competitive_2001,wang_outflow_2010} or Protostellar Merger \citep{bonnell_formation_1998} scenarios of massive star formation. However, if such growth occurs in the protostellar phase, which is expected to last approximately one local free-fall time of a core, then the issue is more subtle. As mentioned, the fiducial Turbulent Core Model predicts about a factor of two growth in the total protostellar core system mass during the protostellar phase, although this would be reduced if outflow feedback strengths during the evolution, which is expected. If the protostellar core was to change its mass by a larger factor, $\gtrsim10$, during the protostellar phase (by either accretion of ambient clump gas or already bound prestellar cores), i.e., so that there was no longer significant correspondence between initial prestellar core mass and final system mass, then this would be the regime of Competitive Accretion. The Protostellar Merger regime would apply when the mass growth of the main protostar is controlled mostly by mergers with other protostars (i.e., near stellar density objects). Thus observations of variations in the CMF, in particular our result that it appears to be more top heavy in higher mass surface density environments, are, on their own, not sufficient to discriminate between these different massive star formation scenarios.

\section{Discussion and Conclusions}\label{S:conclusion}

We have measured the CMF via a sample of 222 dense cores identified via a standardized dendrogram algorithm in mm continuum images of 28 massive ($\gtrsim 500\:M_\odot$) Galactic clumps within 3.5 kpc of the Sun, including accounting for flux and number recovery incompleteness of the cores. We note that while these 28 clumps are not a complete sample of such objects within this region of the Galaxy, we consider that they provide a representative sampling of the conditions in the central regions of massive protoclusters.

We studied where power-law behavior is likely to begin in the CMF through analyzing the behavior of the the corrected Akaike Information Criterion and Anderson-Darling and Kolmogorov-Smirnov statistics, and found that a ``break'' in the power-law behavior of the true CMF is indicated between $\sim$5 to $15 \ M_{\odot}$. We find that a single power-law becomes an increasingly probable descriptor of the CMF above this range and report a best-fit $\alpha = 0.94 \pm 0.08$ for $M \gtrsim 5 \: M_{\odot}$.  

The power-law index we derived is significantly shallower than the standard Salpeter IMF where $\alpha = 1.35$ \citep[see also][]{kroupa_variation_2001} and many previous studies of the local CMF. \citet[see also \citealt{rodon_fragmentation_2012}]{beuther_fragmentation_2004} found a CMF derived from 1.3~mm emission cores in IRAS~19410+2336 ($d\sim2\:$kpc) with $\alpha\simeq1.5\pm0.3$ based on a sample of 24 sources ranging in mass from $\sim2-25\:M_\odot$.  \citet{ohashi_dense_2016} found $\alpha \simeq 1.6\pm0.7$ for the pre-stellar CMF derived from 48 cores with the masses ranging from $1.5-22\:M_\odot$ in IRDC G14.225-0.506 ($d=2\:$kpc) as identified through 3~mm continuum emission.  \citet{shimajiri_probing_2019} reported a peak at $\sim 10 \ M_{\odot}$ in the 3.1 mm continuum-derived CMF of 26 cores in the NGC 6334 filament ($d \sim 1.7$ kpc) observed by ALMA; uncertainties on this result are large due to the small sample size and a limited mass range.
Our analysis of a sample of 222 cores with a large dynamic range of masses represents, to our knowledge, the first ALMA-derived, statistically robust detection of a break in the CMF for sources with average distances $>1\:$kpc. However, while the evidence for a break in the CMF is very strong, is not yet possible with these data to derive its precise location, which we estimate to be somewhere in the range from $\sim5$ to 15$\:M_\odot$.

We also observed a larger proportion of high-mass cores than the IMF and local CMF might indicate.  This shallowness and large high-mass population of cores is consistent with other recent studies of the non-local ($d>1\:$kpc) CMF.  \citet{motte_unexpectedly_2018} studied the 1.3 mm continuum derived CMF in from a sample of 105 cores in W43-MM1 at $d= 5.5$ kpc and derived $\alpha = 0.96 \pm 0.13$ for cores ranging in mass from 1.6 to 100 $M_{\odot}$; their results also showed a larger number of high-mass cores than would typically be expected.  \citet{sanhueza_alma_2019} found $\alpha = 1.07 \pm 0.09$ for $M \geq 0.6 M_{\odot}$ from a sample of 294 pre and protostellar cores identified using the dendrogram algorithim in 12 IRDC clumps ($d$ ranging from 2.9 to 5.4 kpc) observed at 1.3 mm by ALMA.  Similarly, \citet{zhang_fragmentation_2015} identified 38 cores through 1.3~mm emission in the Infrared Dark Cloud (IRDC) G28.34 P1 clump ($d\simeq5\:$kpc) with ALMA and concluded that the population of lower-mass ($\sim1-2\:M_\odot$) cores was lower than expected by predictions derived from the Salpeter mass function and expected star formation efficiency rates. Our Paper II results also indicated a relatively shallow CMF with $\alpha = 0.86 \pm 0.11$ in a larger IRDC-based sample of 107 cores.

However, in most cases the precise values of $\alpha$ are difficult to compare between regions due to differences in fitting methods and large uncertainties in the core mass estimates.  Several other recent studies of the CMF have applied the same MLE and KS-statistic minimizing method proposed by \citet{clauset_power-law_2009} that we use here.  \citet{swift_discerning_2010} applied this and similar techniques to 14 data sets from the literature and found typical values of $M_{\rm{min}}$ for regions with limiting masses $\lesssim 1$ $M_{\odot}$ of $\sim 2 $ $M_{\odot}$.  \citet{olmi_shape_2013} also applied the combined MLE and KS method and derived $\alpha = 1.20 \pm 0.15$ for $M_{\rm{min}}$ $=7.3 \: M_{\odot}$ in a sample of fields observed as part of the Hi-GAL survey ($d=3-8$ kpc).  \citet{lu_alma_2020} derived $\alpha = (0.83 - 1.07)$ for $M \geq 5.9 \  M_{\odot}$ using the same fitting method from a sample of cores identified using the dendrogram algorithim in four massive molecular clouds in the Central Molecular Zone ($d\sim8.2$ kpc) observed by ALMA; they attributed this value of $M_{\rm{min}}$ to the confusion limit within the highly clustered regions they observed, and not to a peak or break.  

Although more relevant, these results are still difficult to compare directly to ours due to variations in core identification and fitting methods and our corrections for flux and number recovery.  By following the methods of Papers I and II in this work, we are able to directly compare our results.  We found that the CMF from this sample of massive clumps has a significantly larger population of high-mass cores than the CMFs derived from the protocluster G286 in Paper I and IRDC clumps in Paper II. Furthermore, the ``true'' CMF we have derived shows clear evidence for a break that occurs at a mass in the range $\sim5$ to $15\:M_\odot$, depending on how it is assessed: a single power law fit is judged to be acceptable at the high mass end above about $5\:M_\odot$, while a double PPL fit is preferred when the fitting range is extended down to $M_{\rm min}\sim 1\:M_\odot$ and the best fit double PPL has its break at about $15\:M_\odot$.

Such a break in the power-law behavior of the CMF between about 5 and $15\:M_{\odot}$ can be compared to equivalent break-like features in the stellar IMF. A single Salpeter power law fit to the IMF appears to be valid down to stellar masses $\sim0.5$ to $2\:M_\odot$ \citep[e.g.][]{bastian_universal_2010}. Accounting for stellar multiplicity, we may then associate such a break mass with a scale of $\sim 1.5\:M_\odot$ for the mass of stars that form from a core. Comparing with our range of CMF break scales, implies a range star formation efficiencies of $\epsilon_{\rm core} \sim 0.1$ to 0.3. Such values are similar to, although slightly lower than, those that have been invoked previously to match observed CMFs with the stellar IMF \citep[e.g.,][]{alves_mass_2007,2010A&A...518L.102A, konyves_census_2015}. If higher values of star formation efficiency are applicable, e.g., $\epsilon_{\rm core}\sim0.3$ to 0.5 that are predicted by theoretical models of star formation feedback \citep{tanaka_impact_2017}
then our results would indicate that the resulting IMF in these central regions of massive protoclusters will be top-heavy. Higher resolution studies are needed to examine how the derived CMF, especially its apparent break scale, depends on resolution.

We note that the clumps observed by the ALMAGAL project were chosen to be especially dense and high-mass, and so, as discussed above, a true variation in the CMF and IMF in these regions would be quite plausible. Still, the following caveats should be noted. The core masses we report have significant uncertainties due to no correction being made for the local backgrounds within which the cores are embedded, and also from a lack of temperature information on the scales of the cores. If, for example, brighter cores are systematically warmer than fainter cores, then their masses would be systematically overestimated in our analysis. It is also likely that a large fraction of the cores observed here are protostellar, i.e., already in the process of collapse and with strong temperature gradients due to internal heating. However, in the absence of detailed spectral energy distribution fitting and other methods of protostar characterization, it is not known how significant an effect this could have on the CMF derived here.   

Future work should include higher sensitivity and higher angular resolution follow-up observations to confirm our results and further probe the distribution of low-mass cores and our observed probable break in the CMF.  Improved temperature measurements of individual cores are also needed to improve their mass measurements and evolutionary stage.

\begin{acknowledgements}
TJO acknowledges support from a VICO undergraduate research fellowship. GC acknowledges support from a Chalmers Cosmic Origins postdoctoral fellowship. JCT acknowledges support from ERC Advanced Grant MSTAR. This paper makes use of the following ALMA data: ADS/JAO.ALMA$\#$2019.1.00195.L. ALMA is a partnership of ESO (representing its member states), NSF (USA) and NINS (Japan), together with NRC (Canada), MOST and ASIAA (Taiwan), and KASI (Republic of Korea), in cooperation with the Republic of Chile. The Joint ALMA Observatory is operated by ESO, AUI/NRAO and NAOJ.  The National Radio Astronomy Observatory is a facility of the National Science Foundation operated under cooperative agreement by Associated Universities, Inc.
\end{acknowledgements}

\bibliographystyle{aasjournal}
\bibliography{refs.bib}

\appendix
\startlongtable
\begin{deluxetable*}{ccccccccc}
\tabletypesize{\scriptsize}
\tablecaption{Estimated physical parameters for cores \label{tab:cores} }
\tablehead{
  \colhead{ID} &  \colhead{$\ell$} & \colhead{$b$} & \colhead{$I_{\rm peak}$} &  \colhead{$S_{\nu}$} &  \colhead{$M_{\rm c,raw}$}  &  \colhead{$M_{c}$}  &    \colhead{$R_{c}$} &  \colhead{$\Sigma_{c}$}  \\
 \colhead{}   & \colhead{($\degr$)}  & \colhead{($\degr$)} & \colhead{(mJy $\rm beam^{-1}$)} &  \colhead{(mJy)} &  \colhead{$(M_{\odot})$} &  \colhead{$(M_{\odot})$}  & \colhead{(0.01 pc)} &   \colhead{(g $\rm cm^{-2}$)}   \\
 \vspace{-0.4cm}
 }
\startdata
G18.30c1 &   18.302391 & -0.391211 &   25.55 &  12.26 &   44.84 &   54.77 &    2.67 &   4.15 \\\   G18.30c2 &   18.302589 & -0.388914 &   12.06 &   4.39 &   35.91 &   59.85 &    4.00 &   1.49 \\\   G18.30c3 &   18.303842 & -0.388483 &    6.86 &   3.75 &   16.92 &   28.89 &    2.97 &   1.27 \\\   G18.30c4 &   18.299457 & -0.388150 &    4.65 &   3.66 &    9.76 &   16.71 &    2.28 &   1.24 \\\   G18.30c5 &   18.304539 & -0.386941 &    2.44 &   2.06 &    0.67 &    1.16 &    0.80 &   0.70 \\\   G24.52c1 &   24.524393 & -0.141540 &    1.52 &   1.32 &    0.30 &    0.52 &    0.66 &   0.45 \\\   G24.52c2 &   24.524888 & -0.140346 &    1.77 &   1.55 &    0.54 &    0.93 &    0.82 &   0.53 \\\   G24.52c3 &   24.525191 & -0.139414 &    8.25 &   6.84 &    2.85 &    3.43 &    0.90 &   2.32 \\\   G24.52c4 &   24.524764 & -0.139391 &    7.37 &   6.44 &    2.23 &    2.72 &    0.82 &   2.18 \\\   G24.52c5 &   24.525508 & -0.138882 &    4.12 &   3.64 &    1.26 &    1.82 &    0.82 &   1.23 \\\  G294.51c1 &  294.512906 & -1.623009 &   24.08 &   7.17 &   13.33 &   18.08 &    1.91 &   2.43 \\\  G294.51c2 &  294.511630 & -1.622002 &    8.11 &   7.82 &    0.75 &    0.99 &    0.43 &   2.65 \\\  G294.51c3 &  294.513179 & -1.621769 &    2.76 &   2.43 &    1.01 &    1.74 &    0.90 &   0.82 \\\  G294.51c4 &  294.511100 & -1.621443 &   20.66 &  18.79 &    2.48 &    2.77 &    0.51 &   6.36 \\\  G294.51c5 &  294.511602 & -1.621178 &   33.85 &  22.86 &   13.18 &   14.43 &    1.06 &   7.74 \\\  G294.51c6 &  294.512701 & -1.620449 &    4.68 &   4.24 &    0.56 &    0.91 &    0.51 &   1.44 \\\  G305.13c1 &  305.140887 &  0.065083 &    7.80 &   5.88 &    4.77 &    7.46 &    1.26 &   1.99 \\\  G305.13c2 &  305.138373 &  0.065771 &    6.95 &   4.38 &    7.68 &   12.95 &    1.85 &   1.48 \\\  G305.13c3 &  305.138148 &  0.067147 &   13.27 &   8.31 &   31.17 &   43.16 &    2.71 &   2.81 \\\  G305.13c4 &  305.134998 &  0.067627 &    4.48 &   3.06 &    5.37 &    9.30 &    1.85 &   1.04 \\\  G305.13c5 &  305.137420 &  0.067646 &    6.29 &   5.81 &    2.35 &    3.70 &    0.89 &   1.97 \\\  G305.13c6 &  305.137164 &  0.068657 &   41.76 &  14.15 &   64.18 &   77.15 &    2.98 &   4.79 \\\  G305.13c7 &  305.138168 &  0.069092 &   23.89 &   9.26 &   49.23 &   65.55 &    3.22 &   3.13 \\\  G305.13c8 &  305.136302 &  0.069199 &   10.64 &   7.86 &    9.55 &   13.50 &    1.54 &   2.66 \\\  G305.13c9 &  305.135379 &  0.070660 &    6.55 &   4.49 &    6.07 &   10.18 &    1.62 &   1.52 \\\ G305.13c10 &  305.135288 &  0.071246 &    5.43 &   4.37 &    4.84 &    8.16 &    1.47 &   1.48 \\\  G309.98c1 &  309.977658 &  0.546082 &    3.59 &   3.09 &    1.47 &    2.50 &    0.97 &   1.05 \\\  G309.98c2 &  309.979869 &  0.546699 &    4.65 &   2.86 &   12.98 &   22.19 &    2.98 &   0.97 \\\  G309.98c3 &  309.980955 &  0.547285 &    3.95 &   3.45 &    1.56 &    2.59 &    0.94 &   1.17 \\\  G309.98c4 &  309.981599 &  0.547832 &    6.10 &   4.49 &    7.74 &   11.97 &    1.84 &   1.52 \\\  G309.98c5 &  309.981176 &  0.548421 &    4.26 &   3.76 &    1.80 &    2.92 &    0.97 &   1.27 \\\  G309.99c1 &  309.990712 &  0.510488 &    4.85 &   2.94 &    3.98 &    6.83 &    1.63 &   1.00 \\\  G309.99c2 &  309.990677 &  0.514099 &   18.25 &   5.40 &   26.04 &   38.82 &    3.07 &   1.83 \\\  G314.21c1 &  314.217277 &  0.251344 &    1.64 &   1.42 &    0.50 &    0.87 &    0.83 &   0.48 \\\  G314.21c2 &  314.218448 &  0.252603 &    6.80 &   2.17 &    7.92 &   13.54 &    2.67 &   0.73 \\\  G314.21c3 &  314.219760 &  0.254121 &    2.89 &   1.76 &    1.00 &    1.73 &    1.05 &   0.60 \\\  G314.21c4 &  314.220232 &  0.254196 &    2.22 &   1.75 &    1.07 &    1.86 &    1.09 &   0.59 \\\  G314.21c5 &  314.220403 &  0.254509 &    2.10 &   1.82 &    0.31 &    0.53 &    0.57 &   0.62 \\\  G314.21c6 &  314.218625 &  0.255304 &    1.29 &   1.13 &    0.28 &    0.49 &    0.70 &   0.38 \\\  G314.21c7 &  314.218070 &  0.255543 &    1.79 &   1.32 &    0.68 &    1.18 &    1.00 &   0.45 \\\  G314.21c8 &  314.221200 &  0.257851 &    4.30 &   3.75 &    0.47 &    0.70 &    0.49 &   1.27 \\\  G314.22c1 &  314.219965 &  0.270431 &    8.17 &   6.22 &    6.46 &   10.56 &    1.43 &   2.10 \\\  G314.22c2 &  314.219357 &  0.270785 &    5.02 &   4.11 &    2.77 &    4.79 &    1.15 &   1.39 \\\  G314.22c3 &  314.222650 &  0.270868 &    4.92 &   3.54 &    1.23 &    2.12 &    0.82 &   1.20 \\\  G314.22c4 &  314.219931 &  0.270995 &    6.02 &   5.16 &    1.51 &    2.57 &    0.76 &   1.75 \\\  G314.22c5 &  314.219621 &  0.271320 &    4.90 &   4.22 &    2.02 &    3.49 &    0.97 &   1.43 \\\  G314.22c6 &  314.219332 &  0.271548 &    4.41 &   4.10 &    0.47 &    0.82 &    0.48 &   1.39 \\\  G314.22c7 &  314.219095 &  0.271720 &    4.73 &   4.03 &    0.93 &    1.61 &    0.67 &   1.36 \\\  G314.22c8 &  314.220142 &  0.272014 &   13.44 &   7.81 &    9.14 &   14.00 &    1.51 &   2.64 \\\  G314.22c9 &  314.220854 &  0.272064 &   17.49 &  12.76 &   18.23 &   23.45 &    1.67 &   4.32 \\\ G314.22c10 &  314.218748 &  0.272094 &    4.76 &   4.36 &    0.58 &    1.00 &    0.51 &   1.48 \\\ G314.22c11 &  314.219422 &  0.272519 &   13.00 &  12.23 &    1.63 &    2.12 &    0.51 &   4.14 \\\ G314.22c12 &  314.220481 &  0.272622 &   15.43 &  11.81 &    7.23 &    9.53 &    1.09 &   4.00 \\\ G314.22c13 &  314.219826 &  0.272665 &   34.62 &  17.57 &   21.36 &   25.66 &    1.54 &   5.95 \\\ G314.22c14 &  314.221166 &  0.272786 &   13.76 &  12.78 &    2.16 &    2.77 &    0.57 &   4.33 \\\ G314.22c15 &  314.220872 &  0.272947 &   16.26 &  13.81 &    8.34 &   10.49 &    1.09 &   4.67 \\\  G316.58c1 &  316.586428 & -0.809995 &    3.47 &   3.15 &    1.43 &    2.31 &    0.94 &   1.07 \\\  G316.58c2 &  316.587046 & -0.809038 &   14.54 &   6.18 &   36.37 &   46.67 &    3.39 &   2.09 \\\  G316.58c3 &  316.587838 & -0.808615 &    3.23 &   2.95 &    1.07 &    1.76 &    0.84 &   1.00 \\\  G316.58c4 &  316.586661 & -0.808321 &    4.38 &   3.98 &    1.09 &    1.63 &    0.73 &   1.35 \\\  G316.58c5 &  316.588566 & -0.808015 &    2.13 &   1.62 &    1.22 &    2.11 &    1.21 &   0.55 \\\  G316.58c6 &  316.587024 & -0.807899 &    5.65 &   4.57 &    4.98 &    7.11 &    1.46 &   1.55 \\\  G316.80c1 &  316.800718 & -0.060433 &   22.54 &  16.98 &   15.21 &   22.87 &    1.32 &   5.75 \\\  G316.80c2 &  316.800539 & -0.059214 &   14.67 &  10.43 &   23.03 &   39.20 &    2.08 &   3.53 \\\  G316.80c3 &  316.798309 & -0.058087 &   12.97 &   8.52 &    9.13 &   15.77 &    1.45 &   2.88 \\\  G316.80c4 &  316.802781 & -0.058070 &   17.67 &  13.59 &   20.01 &   32.23 &    1.70 &   4.60 \\\  G316.80c5 &  316.800743 & -0.057742 &   60.35 &  32.61 &   88.70 &  108.64 &    2.31 &  11.04 \\\  G316.80c6 &  316.798995 & -0.056739 &   18.13 &  17.04 &    5.18 &    7.78 &    0.77 &   5.77 \\\  G316.80c7 &  316.799733 & -0.056498 &   46.27 &  30.14 &   74.27 &   92.33 &    2.20 &  10.20 \\\  G316.80c8 &  316.799082 & -0.055923 &   22.46 &  21.25 &    6.12 &    8.46 &    0.75 &   7.19 \\\  G316.80c9 &  316.798188 & -0.055040 &   33.62 &  27.35 &   35.02 &   44.54 &    1.58 &   9.26 \\\ G316.80c10 &  316.797570 & -0.054580 &   34.60 &  30.12 &   24.58 &   30.56 &    1.26 &  10.19 \\\ G316.80c11 &  316.796076 & -0.053756 &   20.72 &  14.71 &   13.18 &   20.75 &    1.32 &   4.98 \\\ G316.80c12 &  316.797284 & -0.053410 &   20.16 &  17.95 &    5.17 &    7.62 &    0.75 &   6.08 \\\  G316.81c1 &  316.812073 & -0.058407 &    7.90 &   7.25 &    3.02 &    5.20 &    0.90 &   2.45 \\\  G316.81c2 &  316.810106 & -0.057976 &   70.88 &  43.02 &   54.38 &   61.63 &    1.57 &  14.56 \\\  G316.81c3 &  316.810894 & -0.057244 &  100.34 &  48.34 &   92.04 &  102.82 &    1.93 &  16.36 \\\  G317.40c1 &  317.407517 &  0.107427 &    9.45 &   6.05 &    5.84 &    9.29 &    1.37 &   2.05 \\\  G317.40c2 &  317.407456 &  0.108188 &    9.84 &   6.21 &    9.81 &   15.49 &    1.76 &   2.10 \\\  G317.40c3 &  317.408421 &  0.108527 &   18.89 &   6.94 &   13.90 &   21.19 &    1.98 &   2.35 \\\  G317.40c4 &  317.407044 &  0.108983 &   16.55 &   7.80 &   16.29 &   23.85 &    2.02 &   2.64 \\\  G317.40c5 &  317.411358 &  0.109336 &    4.05 &   3.46 &    1.21 &    2.10 &    0.83 &   1.17 \\\  G317.40c6 &  317.412304 &  0.109425 &    5.24 &   4.54 &    1.75 &    2.97 &    0.87 &   1.54 \\\  G317.40c7 &  317.408454 &  0.109620 &   22.94 &   8.37 &   38.36 &   54.66 &    2.99 &   2.83 \\\  G317.40c8 &  317.407727 &  0.110473 &    5.56 &   4.57 &    5.54 &    9.39 &    1.54 &   1.55 \\\  G317.40c9 &  317.406947 &  0.112833 &    4.64 &   3.64 &    2.05 &    3.53 &    1.05 &   1.23 \\\  G319.86c1 &  319.866187 &  0.784905 &    2.88 &   1.75 &    1.75 &    3.03 &    1.40 &   0.59 \\\  G319.86c2 &  319.866516 &  0.785668 &    1.94 &   1.29 &    2.43 &    4.21 &    1.92 &   0.44 \\\  G319.86c3 &  319.870020 &  0.785786 &    3.81 &   2.44 &    4.26 &    7.12 &    1.85 &   0.82 \\\  G319.86c4 &  319.868780 &  0.786121 &    5.01 &   2.12 &    7.48 &   12.77 &    2.63 &   0.72 \\\  G319.86c5 &  319.864519 &  0.786161 &    6.89 &   2.74 &    5.41 &    8.79 &    1.96 &   0.93 \\\  G319.86c6 &  319.863747 &  0.786554 &    4.31 &   2.01 &    4.75 &    8.14 &    2.15 &   0.68 \\\  G319.86c7 &  319.866407 &  0.786825 &    1.52 &   1.17 &    0.46 &    0.80 &    0.87 &   0.40 \\\  G319.86c8 &  319.867597 &  0.787110 &    1.17 &   1.03 &    0.27 &    0.47 &    0.71 &   0.35 \\\  G319.86c9 &  319.866731 &  0.787660 &    1.69 &   1.24 &    1.69 &    2.92 &    1.63 &   0.42 \\\ G319.86c10 &  319.865327 &  0.787898 &    2.29 &   1.46 &    3.85 &    6.68 &    2.27 &   0.49 \\\ G319.86c11 &  319.869534 &  0.789316 &    2.93 &   2.32 &    0.91 &    1.53 &    0.87 &   0.78 \\\ G319.86c12 &  319.870033 &  0.789482 &    4.43 &   3.13 &    2.15 &    3.37 &    1.16 &   1.06 \\\  G320.22c1 &  320.228208 &  0.871500 &    3.78 &   2.27 &    3.49 &    5.66 &    1.73 &   0.77 \\\  G320.22c2 &  320.227613 &  0.872132 &    1.37 &   1.17 &    1.26 &    2.19 &    1.45 &   0.40 \\\  G320.22c3 &  320.228575 &  0.872265 &    5.31 &   2.60 &    5.17 &    8.08 &    1.97 &   0.88 \\\  G320.22c4 &  320.226638 &  0.872345 &    0.99 &   0.84 &    0.36 &    0.62 &    0.91 &   0.29 \\\  G320.22c5 &  320.227159 &  0.872603 &    1.16 &   1.02 &    0.23 &    0.41 &    0.67 &   0.35 \\\  G320.24c1 &  320.246683 & -0.296828 &    5.05 &   4.70 &    0.08 &    0.12 &    0.18 &   1.59 \\\  G320.24c2 &  320.246490 & -0.295853 &    9.48 &   7.50 &    0.84 &    1.10 &    0.47 &   2.54 \\\  G320.24c3 &  320.245650 & -0.295431 &   26.14 &  12.02 &    3.70 &    4.36 &    0.78 &   4.07 \\\  G320.24c4 &  320.248669 & -0.294649 &    2.27 &   1.93 &    0.04 &    0.07 &    0.21 &   0.65 \\\  G320.24c5 &  320.250530 & -0.294527 &    4.13 &   3.62 &    0.09 &    0.15 &    0.22 &   1.22 \\\  G320.24c6 &  320.247912 & -0.294307 &    2.39 &   1.86 &    0.09 &    0.15 &    0.30 &   0.63 \\\  G320.24c7 &  320.244955 & -0.294242 &   12.10 &   6.10 &    1.26 &    1.77 &    0.64 &   2.06 \\\  G320.24c8 &  320.246137 & -0.294037 &   15.64 &   9.26 &    2.75 &    3.40 &    0.76 &   3.13 \\\  G326.34c1 &  326.340984 &  0.504252 &    4.20 &   2.34 &    2.40 &    4.14 &    1.42 &   0.79 \\\  G326.34c2 &  326.340939 &  0.504686 &    2.41 &   1.92 &    0.46 &    0.79 &    0.68 &   0.65 \\\  G326.34c3 &  326.340918 &  0.505075 &    1.98 &   1.84 &    0.15 &    0.26 &    0.40 &   0.62 \\\  G326.34c4 &  326.341014 &  0.505492 &    7.84 &   3.50 &    3.86 &    6.23 &    1.47 &   1.18 \\\  G327.39c1 &  327.394501 &  0.196555 &    9.10 &   3.83 &    7.33 &   11.69 &    1.94 &   1.29 \\\  G327.39c2 &  327.393753 &  0.197325 &    5.19 &   2.44 &    3.46 &    5.95 &    1.66 &   0.83 \\\  G327.39c3 &  327.393028 &  0.197911 &    9.01 &   2.80 &    4.64 &    7.91 &    1.80 &   0.95 \\\  G327.39c4 &  327.392052 &  0.198893 &    8.04 &   6.08 &    4.97 &    6.74 &    1.26 &   2.06 \\\  G327.39c5 &  327.391444 &  0.199291 &   10.80 &   9.33 &    1.80 &    2.19 &    0.62 &   3.16 \\\  G327.39c6 &  327.391756 &  0.199403 &    8.65 &   7.54 &    2.98 &    3.78 &    0.88 &   2.55 \\\  G327.39c7 &  327.391090 &  0.199533 &   26.22 &  14.46 &   11.81 &   13.37 &    1.26 &   4.89 \\\  G327.39c8 &  327.392203 &  0.199630 &    5.77 &   5.53 &    0.83 &    1.16 &    0.54 &   1.87 \\\  G327.39c9 &  327.392759 &  0.199935 &    1.66 &   1.51 &    0.26 &    0.46 &    0.59 &   0.51 \\\ G327.39c10 &  327.393136 &  0.200261 &    2.11 &   1.58 &    0.61 &    1.06 &    0.87 &   0.54 \\\ G327.39c11 &  327.392087 &  0.200533 &    2.16 &   1.63 &    0.59 &    1.02 &    0.84 &   0.55 \\\ G327.39c12 &  327.391387 &  0.200548 &    2.89 &   2.10 &    1.71 &    2.97 &    1.26 &   0.71 \\\ G327.39c13 &  327.391249 &  0.201118 &    1.83 &   1.57 &    0.32 &    0.55 &    0.63 &   0.53 \\\ G327.39c14 &  327.389139 &  0.202845 &    5.47 &   4.52 &    0.87 &    1.32 &    0.62 &   1.53 \\\  G341.92c1 &  341.931423 & -0.172863 &    3.27 &   2.80 &    1.58 &    2.65 &    1.05 &   0.95 \\\  G341.92c2 &  341.930483 & -0.170696 &    1.91 &   1.50 &    3.02 &    5.25 &    1.98 &   0.51 \\\  G341.92c3 &  341.929905 & -0.169085 &    5.14 &   2.18 &   21.91 &   37.71 &    4.44 &   0.74 \\\  G341.92c4 &  341.930386 & -0.168042 &    2.02 &   1.52 &    1.10 &    1.91 &    1.19 &   0.52 \\\  G341.92c5 &  341.928756 & -0.167222 &    3.52 &   1.79 &    4.65 &    8.06 &    2.26 &   0.61 \\\  G341.92c6 &  341.927964 & -0.165168 &    6.62 &   4.12 &    4.38 &    6.55 &    1.44 &   1.39 \\\  G341.94c1 &  341.940847 & -0.170856 &    9.45 &   8.33 &    3.96 &    5.75 &    0.96 &   2.82 \\\  G341.94c2 &  341.945718 & -0.167906 &   13.07 &  10.19 &    8.30 &   11.18 &    1.26 &   3.45 \\\  G341.94c3 &  341.942391 & -0.167692 &    6.39 &   3.96 &    7.88 &   13.58 &    1.97 &   1.34 \\\  G341.94c4 &  341.945315 & -0.167608 &    8.12 &   7.30 &    2.81 &    4.27 &    0.87 &   2.47 \\\  G341.94c5 &  341.943153 & -0.166998 &   16.61 &  10.42 &   10.84 &   14.49 &    1.43 &   3.53 \\\  G341.94c6 &  341.943379 & -0.166437 &   15.54 &   9.78 &   17.03 &   23.29 &    1.85 &   3.31 \\\  G341.94c7 &  341.943986 & -0.166390 &    6.78 &   6.27 &    2.27 &    3.62 &    0.84 &   2.12 \\\  G341.94c8 &  341.941697 & -0.165969 &    3.84 &   3.50 &    1.50 &    2.60 &    0.92 &   1.19 \\\  G341.94c9 &  341.941294 & -0.165622 &    3.90 &   3.62 &    1.72 &    2.97 &    0.96 &   1.23 \\\ G341.94c10 &  341.943269 & -0.165196 &   35.98 &  17.39 &   46.41 &   55.02 &    2.28 &   5.89 \\\ G341.94c11 &  341.942768 & -0.164775 &   13.72 &  12.23 &    4.98 &    6.33 &    0.89 &   4.14 \\\ G341.94c12 &  341.943839 & -0.164708 &    5.14 &   4.52 &    1.84 &    3.14 &    0.89 &   1.53 \\\ G341.94c13 &  341.942398 & -0.164502 &   15.62 &  13.13 &    6.53 &    8.16 &    0.99 &   4.44 \\\ G341.94c14 &  341.940988 & -0.163290 &    4.11 &   3.45 &    1.95 &    3.37 &    1.05 &   1.17 \\\  G343.23c1 &  343.239216 & -0.713103 &    4.26 &   2.12 &    5.56 &    9.35 &    2.27 &   0.72 \\\  G343.23c2 &  343.239358 & -0.712574 &    1.23 &   1.11 &    0.27 &    0.47 &    0.69 &   0.37 \\\  G343.23c3 &  343.240124 & -0.712289 &    1.10 &   0.96 &    0.42 &    0.73 &    0.92 &   0.33 \\\  G343.52c1 &  343.520823 & -0.520052 &    2.50 &   2.08 &    0.34 &    0.59 &    0.56 &   0.70 \\\  G343.52c2 &  343.521149 & -0.517135 &   10.71 &   3.28 &   15.88 &   26.63 &    3.08 &   1.11 \\\  G343.52c3 &  343.522838 & -0.516932 &    1.72 &   1.50 &    0.22 &    0.38 &    0.54 &   0.51 \\\  G343.52c4 &  343.522687 & -0.514724 &    3.21 &   2.43 &    0.50 &    0.87 &    0.64 &   0.82 \\\  G343.90c1 &  343.903898 & -0.671102 &   12.68 &   4.11 &   12.59 &   18.50 &    2.45 &   1.39 \\\  G343.90c2 &  343.904053 & -0.669897 &    2.03 &   1.74 &    0.25 &    0.42 &    0.52 &   0.59 \\\  G344.06c1 &  344.063059 & -0.655100 &    4.54 &   3.59 &    1.50 &    2.23 &    0.90 &   1.21 \\\  G344.06c2 &  344.064262 & -0.652653 &    8.05 &   3.32 &    4.71 &    7.19 &    1.66 &   1.12 \\\  G344.06c3 &  344.064614 & -0.651596 &    9.47 &   3.40 &    5.32 &    8.07 &    1.75 &   1.15 \\\  G344.06c4 &  344.061990 & -0.649426 &    1.26 &   1.09 &    0.32 &    0.56 &    0.76 &   0.37 \\\  G344.06c5 &  344.067220 & -0.648939 &    4.85 &   3.29 &    1.33 &    2.04 &    0.89 &   1.11 \\\  G344.06c6 &  344.066422 & -0.648731 &    3.26 &   2.18 &    1.65 &    2.80 &    1.22 &   0.74 \\\  G344.72c1 &  344.725577 & -0.542795 &    1.30 &   1.02 &    0.35 &    0.61 &    0.82 &   0.34 \\\  G344.72c2 &  344.726270 & -0.541603 &    1.00 &   0.87 &    0.16 &    0.27 &    0.59 &   0.30 \\\  G344.72c3 &  344.727568 & -0.541305 &    3.69 &   1.90 &    2.53 &    4.29 &    1.62 &   0.64 \\\  G344.72c4 &  344.726618 & -0.541177 &    3.83 &   2.05 &    2.22 &    3.71 &    1.46 &   0.69 \\\  G345.18c1 &  345.182537 &  1.044677 &    4.68 &   4.18 &    0.41 &    0.70 &    0.44 &   1.42 \\\  G345.18c2 &  345.183986 &  1.044937 &   16.24 &   6.62 &    4.77 &    7.72 &    1.19 &   2.24 \\\  G345.18c3 &  345.180079 &  1.045351 &   49.47 &  24.47 &   26.41 &   30.19 &    1.45 &   8.28 \\\  G345.18c4 &  345.185668 &  1.045806 &    6.25 &   5.60 &    0.74 &    1.24 &    0.51 &   1.90 \\\  G345.18c5 &  345.181755 &  1.045834 &    7.10 &   6.58 &    2.46 &    3.99 &    0.85 &   2.23 \\\  G345.18c6 &  345.179350 &  1.045955 &   25.53 &  18.69 &    5.30 &    6.32 &    0.74 &   6.33 \\\  G345.18c7 &  345.182093 &  1.046482 &    9.48 &   7.66 &    3.23 &    5.02 &    0.91 &   2.59 \\\  G345.18c8 &  345.178209 &  1.047093 &    5.95 &   4.59 &    1.17 &    2.02 &    0.71 &   1.55 \\\  G345.18c9 &  345.181739 &  1.047139 &   10.23 &   7.12 &    3.00 &    4.77 &    0.91 &   2.41 \\\  G345.33c1 &  345.335112 &  1.017886 &    3.87 &   3.03 &    2.15 &    3.73 &    1.18 &   1.02 \\\  G345.33c2 &  345.333409 &  1.019431 &    5.92 &   5.32 &    0.48 &    0.78 &    0.42 &   1.80 \\\  G345.33c3 &  345.334066 &  1.019551 &   49.09 &  31.46 &   12.06 &   13.11 &    0.87 &  10.63 \\\  G345.33c4 &  345.334715 &  1.019938 &    3.89 &   3.17 &    0.91 &    1.57 &    0.75 &   1.07 \\\  G345.33c5 &  345.333795 &  1.020072 &   27.61 &  23.44 &    3.60 &    4.02 &    0.55 &   7.92 \\\  G345.33c6 &  345.333873 &  1.020884 &   12.65 &   6.51 &    3.13 &    4.76 &    0.97 &   2.20 \\\  G345.33c7 &  345.336066 &  1.020928 &    2.77 &   2.13 &    0.61 &    1.06 &    0.75 &   0.72 \\\  G345.33c8 &  345.335983 &  1.022170 &   22.38 &  20.86 &    1.74 &    1.98 &    0.40 &   7.05 \\\  G345.33c9 &  345.336222 &  1.022451 &   27.15 &  23.09 &    4.19 &    4.68 &    0.60 &   7.80 \\\  G345.50c1 &  345.504917 &  0.343285 &   25.99 &  21.46 &    2.97 &    4.54 &    0.52 &   7.27 \\\  G345.50c2 &  345.501894 &  0.343579 &   34.77 &  27.45 &    7.84 &   10.94 &    0.75 &   9.30 \\\  G345.50c3 &  345.504348 &  0.344091 &   21.20 &  18.48 &    4.64 &    7.42 &    0.70 &   6.26 \\\  G345.50c4 &  345.501924 &  0.344337 &   20.20 &  17.55 &    2.58 &    4.19 &    0.54 &   5.94 \\\  G345.50c5 &  345.504303 &  0.344494 &   18.26 &  15.87 &    2.75 &    4.57 &    0.58 &   5.37 \\\  G345.50c6 &  345.507217 &  0.344735 &   19.56 &  17.34 &    3.75 &    6.10 &    0.65 &   5.87 \\\  G345.50c7 &  345.503949 &  0.344781 &   16.94 &  14.51 &    3.01 &    5.10 &    0.64 &   4.91 \\\  G345.50c8 &  345.505905 &  0.344949 &   20.58 &  15.39 &    6.40 &   10.70 &    0.90 &   5.21 \\\  G345.50c9 &  345.501642 &  0.345132 &   19.58 &  15.36 &    3.86 &    6.46 &    0.70 &   5.20 \\\ G345.50c10 &  345.502798 &  0.345669 &   12.19 &  10.29 &    1.69 &    2.93 &    0.57 &   3.48 \\\ G345.50c11 &  345.508943 &  0.345692 &   40.81 &  30.07 &   10.15 &   13.70 &    0.81 &  10.18 \\\ G345.50c12 &  345.508250 &  0.345821 &   21.84 &  18.59 &    2.25 &    3.59 &    0.49 &   6.30 \\\ G345.50c13 &  345.504499 &  0.346099 &   29.20 &  23.01 &    6.78 &   10.10 &    0.76 &   7.79 \\\ G345.50c14 &  345.499556 &  0.346174 &   27.67 &  24.31 &    2.95 &    4.31 &    0.49 &   8.23 \\\ G345.50c15 &  345.508944 &  0.346379 &   36.25 &  26.43 &    9.15 &   12.96 &    0.82 &   8.95 \\\ G345.50c16 &  345.504556 &  0.347022 &   34.14 &  32.33 &    7.00 &    9.21 &    0.65 &  10.95 \\\ G345.50c17 &  345.508621 &  0.347317 &   21.56 &  17.89 &    2.17 &    3.50 &    0.49 &   6.06 \\\ G345.50c18 &  345.509484 &  0.347484 &   33.95 &  28.47 &    4.19 &    5.77 &    0.54 &   9.64 \\\ G345.50c19 &  345.504453 &  0.348168 &  381.73 &  90.05 &  251.84 &  277.10 &    2.34 &  30.50 \\\ G345.50c20 &  345.506759 &  0.348617 &   11.18 &   9.35 &    3.89 &    6.74 &    0.90 &   3.17 \\\ G345.50c21 &  345.501019 &  0.348789 &   21.45 &  14.08 &    9.63 &   16.34 &    1.16 &   4.77 \\\ G345.50c22 &  345.499773 &  0.349220 &   26.95 &  21.70 &    8.45 &   12.86 &    0.87 &   7.35 \\\ G345.50c23 &  345.509444 &  0.349374 &   35.12 &  29.40 &    4.84 &    6.58 &    0.57 &   9.96 \\\ G345.50c24 &  345.505236 &  0.349572 &   10.94 &   9.75 &    1.69 &    2.93 &    0.58 &   3.30 \\\ G345.50c25 &  345.500845 &  0.349653 &   16.01 &  13.28 &    1.72 &    2.95 &    0.50 &   4.50 \\\ G345.50c26 &  345.499750 &  0.349762 &   35.48 &  26.63 &   10.15 &   14.32 &    0.86 &   9.02 \\\ G345.50c27 &  345.499466 &  0.350504 &   60.38 &  40.61 &   13.71 &   16.99 &    0.81 &  13.76 \\\ G345.50c28 &  345.502449 &  0.351674 &   20.23 &  15.99 &    4.43 &    7.35 &    0.74 &   5.42 \\\ G345.50c29 &  345.502469 &  0.352271 &   21.01 &  17.99 &    2.49 &    4.01 &    0.52 &   6.09 \\\ G345.50c30 &  345.501464 &  0.352413 &   52.72 &  35.29 &   16.19 &   20.74 &    0.95 &  11.95 \\\ G345.50c31 &  345.504466 &  0.352745 &   22.75 &  18.84 &    2.94 &    4.67 &    0.55 &   6.38 \\\ G345.50c32 &  345.502712 &  0.352796 &   34.19 &  26.64 &   11.30 &   15.95 &    0.91 &   9.02 \\\ G345.50c33 &  345.504388 &  0.353371 &   44.01 &  34.35 &   11.30 &   14.59 &    0.80 &  11.63 \\\  G346.35c1 &  346.355034 &  0.104920 &    3.86 &   1.39 &    1.52 &    2.55 &    1.46 &   0.47 \\\  G346.35c2 &  346.355044 &  0.105452 &    1.15 &   0.75 &    0.38 &    0.67 &    1.00 &   0.25 \\\  G346.35c3 &  346.354345 &  0.105459 &    1.05 &   0.77 &    0.39 &    0.68 &    1.00 &   0.26
\\
\enddata
\tablecomments{$M_c$ is the mass estimate after flux correction, which
equals the raw, uncorrected mass estimate ($M_{\rm c,raw}$) multiplied
by the value of $f_{\rm flux}$ appropriate for $M_c$.}
\end{deluxetable*}

\end{document}